\documentclass[sigconf]{acmart}
\renewcommand\footnotetextcopyrightpermission[1]{} 
\AtBeginDocument{%
  }

\acmConference[arXiv Preprint]{arXiv Preprint}{April 2026}{None}

\usepackage{multirow}
\usepackage{subcaption}
\usepackage{algorithm}
\usepackage[noend]{algorithmic}

\begin{document}

\title{From Entity Reliability to Clean Feedback: An Entity-Aware Denoising Framework Beyond Interaction-Level Signals}

\author{Ze Liu, Xianquan Wang, Shuochen Liu, Jie Ma, Huibo Xu, Yupeng Han, Kai Zhang}
\authornote{Corresponding author.}
\affiliation{%
  \institution{University of Science and Technology of China \& State Key Laboratory of Cognitive Intelligence}
  \country{China}
}
\email{{liuze2120, kkzhang08}@mail.ustc.edu.cn}

\author{Jun Zhou}
\affiliation{%
  \institution{Ant Group}
  \country{China}
}
\email{jun.zhoujun@antfin.com}

\renewcommand{\shortauthors}{Liu and Wang et al.}

\begin{abstract}
Implicit feedback is central to modern recommender systems but is inherently noisy, often impairing model training and degrading user experience. At scale, such noise can mislead learning processes, reducing both recommendation accuracy and platform value. Existing denoising strategies typically overlook the entity-specific nature of noise while introducing high computational costs and complex hyperparameter tuning. To address these challenges, we propose \textbf{EARD} (\textbf{E}ntity-\textbf{A}ware \textbf{R}eliability-\textbf{D}riven Denoising), a lightweight framework that shifts the focus from interaction-level signals to entity-level reliability. Motivated by the empirical observation that training loss correlates with noise, EARD quantifies user and item reliability via their average training losses as a proxy for reputation, and integrates these entity-level factors with interaction-level confidence. The framework is \textbf{model-agnostic}, \textbf{computationally efficient}, and requires \textbf{only two intuitive hyperparameters}. Extensive experiments across multiple datasets and backbone models demonstrate that EARD yields substantial improvements over state-of-the-art baselines (e.g., up to 27.01\% gain in NDCG@50), while incurring negligible additional computational cost. Comprehensive ablation studies and mechanism analyses further confirm EARD's robustness to hyperparameter choices and its practical scalability. These results highlight the importance of entity-aware reliability modeling for denoising implicit feedback and pave the way for more robust recommendation research. 
\end{abstract}




\maketitle

\section{Introduction}

Recommender systems are essential to modern information platforms, supporting personalized content delivery in e-commerce and social media. These systems increasingly rely on large-scale implicit feedback, such as clicks and viewing histories, to model user preferences. Implicit signals are abundant and require no explicit input, making them well-suited for industrial-scale deployment.

\begin{figure}[tb]
\centering
\begin{subfigure}[t]{1.0\linewidth}
    \centering
    \includegraphics[width=1.0\linewidth, keepaspectratio]{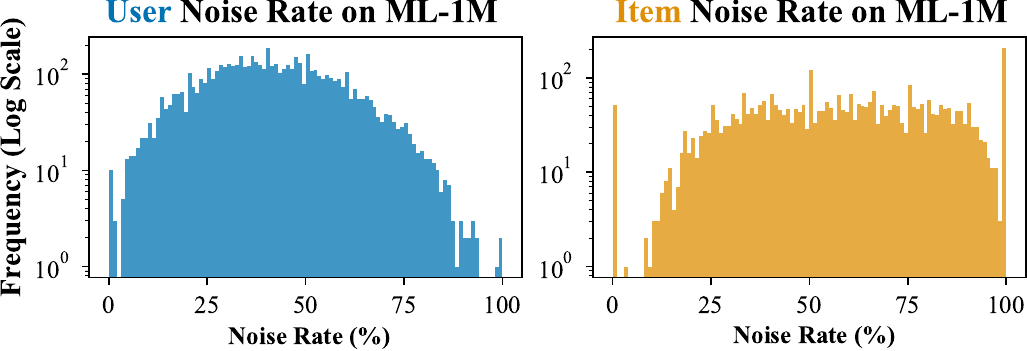}
    \Description{User and item noise rate distributions on ML-1M dataset. Ratings <=3 are labeled as noise.}
    \caption{ }
    \label{fig:ml_1m_noisy_rate_distribution}
\end{subfigure}
\begin{subfigure}[t]{1.0\linewidth}
    \centering
    \includegraphics[width=1.0\linewidth, keepaspectratio]{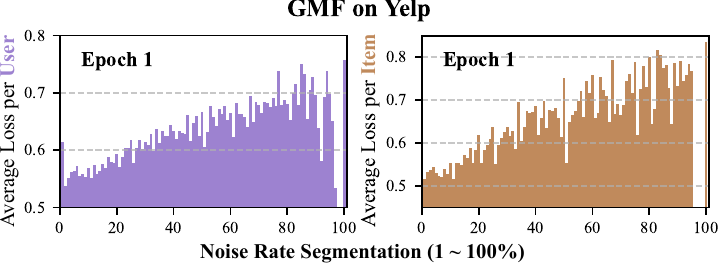}
    \Description{Average loss of user and item groups with different noise levels on Yelp using GMF model.}
    \caption{ }
    \label{fig:user_item_loss_histograms_GMF_Yelp_epoch1}
\end{subfigure}
\vspace{-1.0em}
\caption{(a) User (blue) and item (yellow) noise rate distributions on ML-1M. As in ADT~\cite{ADT}, ratings $\leq$ 3 are labeled as noise. The broad spread shows that entity reliability varies widely.
(b) Average loss of user (purple) and item (brown) groups with different noise levels on Yelp using GMF. Average loss tends to increase with noise rate.}
\label{fig:ml_1m_noisy_rate_distribution_and_user_item_loss_histograms_GMF_Yelp_epoch1}
\end{figure}

However, implicit feedback is inherently noisy, containing both \emph{false positives} (e.g., accidental clicks or clickbait-induced interactions) and \emph{false negatives} (non-interactions that do not imply disinterest, often due to exposure bias). Such noise can misguide model training and degrade recommendation accuracy. To mitigate this, numerous denoising methods have been proposed to address the noise inherent in implicit feedback for recommendation systems. Prevailing strategies typically involve re-weighting training instances based on their loss values~\cite{ADT}, pruning noisy edges from the user-item interaction graph~\cite{RGCF}, or leveraging more complex mechanisms like cross-model disagreement~\cite{DeCA, SGDL}. Recommendation systems rely on massive amounts of data and incur substantial training costs, so any additional denoising mechanism must be lightweight, efficient, and compatible with the main training pipeline without interfering with or slowing down the original process. This raises a fundamental and critical research question: How can we design a denoising framework that is not only effective at identifying noise sources but also practical for large-scale, real-world deployment? Answering this question requires overcoming three major, interrelated challenges that plague existing methods:

\textbf{Lack of Entity-Aware Modeling.} Most existing methods address noise within the vast and sparse user--item interaction space, yet they typically analyze it only at the interaction level. A critical oversight in these approaches is the failure to {decouple} the origins of noise by attributing it distinctly to the user or the item. For instance, noise can stem from the user side, such as from users who are merely `wandering' and whose clicks do not reflect genuine preference. Conversely, it can originate from the item side, where certain items act as `clickbait,' engineered to attract interactions that are not indicative of true interest. The necessity of this decoupled perspective is substantiated by our analysis of user and item noise rates across the ML-1M, Yelp, and AmazonBook datasets. As shown in Figure~\ref{fig:ml_1m_noisy_rate_distribution}, noise rates for both entities span the full 0--100\% range with substantial variability, a pattern confirmed in other datasets (Appendix Figure~\ref{fig:all_noisy_rate_distribution}). These findings clearly invalidate the implicit assumption that all entities are equally reliable and underscore the urgent need for entity-aware denoising approaches. By modeling user and item reliability independently, we can more accurately assess interaction quality and improve the identification and suppression of noisy signals.

\textbf{High Computational Overhead.} To improve denoising effectiveness, several methods, including SGDL~\cite{SGDL}, DeCA~\cite{DeCA}, and BOD~\cite{BOD}, adopt complex optimization strategies such as meta-learning frameworks, second-order gradient computations, and multi-path auxiliary networks. While these approaches can achieve strong performance in small- or medium-scale experiments, they introduce substantial training and inference costs. This overhead becomes prohibitive in industrial-scale scenarios with tens of millions of interactions, limiting their practicality for real-time recommendation systems that demand both efficiency and scalability.

\textbf{Excessive Hyperparameter Complexity.} Many existing methods require multiple hyperparameters to manage loss weighting, sample selection, or optimization. For example, DeCA, UDT, and SGDL introduce three to five core hyperparameters. This increases tuning costs and reduces stability across datasets, hindering generalization. Extensive hyperparameter search is often impractical in real-world scenarios. Prior work ~\cite{TheSimplerTheBetter} shows that simpler models can perform as well as more complex ones under fair evaluation. Effective denoising methods should therefore reduce manual tuning and remain robust under default settings.

To address these challenges, we estimate an entity's noise level using a simple yet effective proxy. Specifically, for each user (or item), we compute its average training loss over all associated interactions in an epoch. This approach is inspired by prior works~\cite{Curriculumlearning, Co-teaching, MentorNet, ADT}, which show that noisy samples tend to incur higher losses. To validate this assumption, we grouped users and items by noise rate bins and computed their average losses during the first epoch of GMF on the Yelp dataset. As shown in Figure~\ref{fig:user_item_loss_histograms_GMF_Yelp_epoch1}, the average loss of entities generally increases with their noise rate. Additional evidence in Appendix~\ref{sec:avgLoss_noiseRate} (Figures~\ref{fig:user_loss_histograms} and \ref{fig:item_loss_histograms}) confirms this trend.

Based on this observation, we propose a lightweight three-stage pipeline to estimate fine-grained confidence scores: (1) compute the average loss for each entity with minimal overhead; (2) weight entity-level scores using interaction-level weights derived from the empirical cumulative distribution function (ECDF) of individual losses; (3) transform losses into reliability scores via a linear mapping parameterized by $\alpha$ and $\beta$. This design captures both entity reliability and interaction uncertainty, while remaining model-agnostic, efficient, and requiring minimal hyperparameter tuning. Our key contributions are as follows:

\noindent\textbf{1) Entity-Aware denoising.} We propose EARD, a novel and effective framework that explicitly incorporates both user and item reliability to better detect noise in implicit feedback.

\noindent\textbf{2) Scalable and efficient design.} Our method computes entity- and interaction-level weights with negligible overhead, enabling application to large-scale industrial datasets.

\noindent\textbf{3) Minimal hyperparameter requirement.} The framework introduces only two intuitive hyperparameters, $\alpha$ and $\beta$, with a smooth and unimodal performance landscape that supports robust tuning.
    
\noindent\textbf{4) Comprehensive empirical validation.} Experiments across multiple datasets and representative models show consistent and significant improvements over state-of-the-art denoising baselines, with lower computational and memory costs.

\section{Related Work}

Denoising implicit feedback is essential for robust recommendation, as it often includes false positives (\textit{e.g.}, accidental clicks) and false negatives (\textit{e.g.}, unobserved preferences due to exposure bias). Early approaches used heuristic rules and statistical reweighting. \textbf{WRMF}~\cite{WRMF} separates interactions into binary preferences and confidence scores based on frequency, optimizing with Alternating Least Squares for scalability and interpretability. \textbf{WBPR}~\cite{WBPR} extends Bayesian Personalized Ranking by assigning popularity-aware sampling to negatives, mitigating sampling bias.

Later work leverages the memory effect in deep learning, where clean samples are learned earlier. \textbf{ADT}~\cite{ADT} uses truncated and reweighted losses to suppress high-loss samples. \textbf{IR}~\cite{IR} iteratively refines labels through self-training. \textbf{SGDL}~\cite{SGDL} adopts a two-phase approach that detects memorized samples, then adjusts loss weights via meta-learning and an LSTM scheduler. \textbf{DCF}~\cite{DCF} smooths loss histories and relabels samples to retain informative ones while filtering noise. \textbf{UDT}~\cite{UDT} models interactions as a two-stage (``willingness'' to ``action'') Markov process and applies hierarchical loss weighting. \textbf{PLD}~\cite{PLD} resamples interactions using user-specific loss distributions that better separate clean from noisy data.

Recent methods address noise using structural signals, multi-view learning, and modality-specific features. \textbf{DRPN}~\cite{DRPN} filters noise by comparing aggregated feedback. \textbf{DeCA}~\cite{DeCA} detects noisy labels via cross-model disagreement. \textbf{RGCF}~\cite{RGCF} prunes noisy edges in graph-based models using similarity scores and multi-view mutual information. \textbf{KRDN}~\cite{KRDN} denoises knowledge graph triples and user-item links through attention masking and graph contrast. \textbf{RocSE}~\cite{RocSE} improves robustness through structural filtering, embedding noise, and contrastive learning. \textbf{SLED}~\cite{SLED} uses local graph patterns in a two-stage unsupervised framework to estimate reliability. \textbf{DA-MRS}~\cite{DA-MRS} handles multimodal noise via cross-modal consistency, noise-aware BPR loss, and contrastive objectives.

Beyond structural and loss-based methods, recent work explores advanced paradigms. \textbf{AutoDenoise}~\cite{AutoDenoise} applies reinforcement learning to select clean samples based on training utility. \textbf{BOD}~\cite{BOD} formulates sample weighting as bi-level optimization guided by validation performance. \textbf{DDRM}~\cite{DDRM} uses diffusion models to generate user and item embeddings from collaborative signals. \textbf{LLaRD}~\cite{LLaRD} employs large language models and side information to infer preferences through chain-of-thought reasoning, enhancing noise detection via information bottleneck regularization.

Building on insights from prior work, we present \textbf{EARD}, an entity-aware framework that integrates user and item reliability into interaction weighting. The design is lightweight, scalable, and model-agnostic, effectively addressing key limitations of existing methods, including the absence of entity-level modeling, high computational cost, and extensive hyperparameter tuning.

\section{Method}

\subsection{Problem Formulation}

\subsubsection{Notation}

Let $\mathcal{U} = \{u_1, u_2, \dots, u_M\}$ and $\mathcal{I} = \{i_1, i_2, \dots, i_N\}$ denote the sets of $M$ users and $N$ items. The set of observed user-item interactions is $\mathcal{O} = \{(u, i) | \text{user } u \text{ interacted with item } i\}$. For training, we use the observed interactions $\mathcal{O}$ as positive instances and randomly sample a set of unobserved interactions $\mathcal{S}$ as negative instances, where $|\mathcal{S}| = k \cdot |\mathcal{O}|$, where $k$ is the negative sampling ratio. The full set of training instances for an epoch is denoted as $\mathcal{D} = \mathcal{O} \cup \mathcal{S}$. Our goal is to learn a prediction function $\hat{y}_{ui} = f(u, i; \Theta)$ that minimizes a loss function, such as Binary Cross-Entropy: $\ell(\hat{y}_{ui}, y_{ui})$.

\begin{figure*}[!t]
\centering
\includegraphics[width=1.0\linewidth]{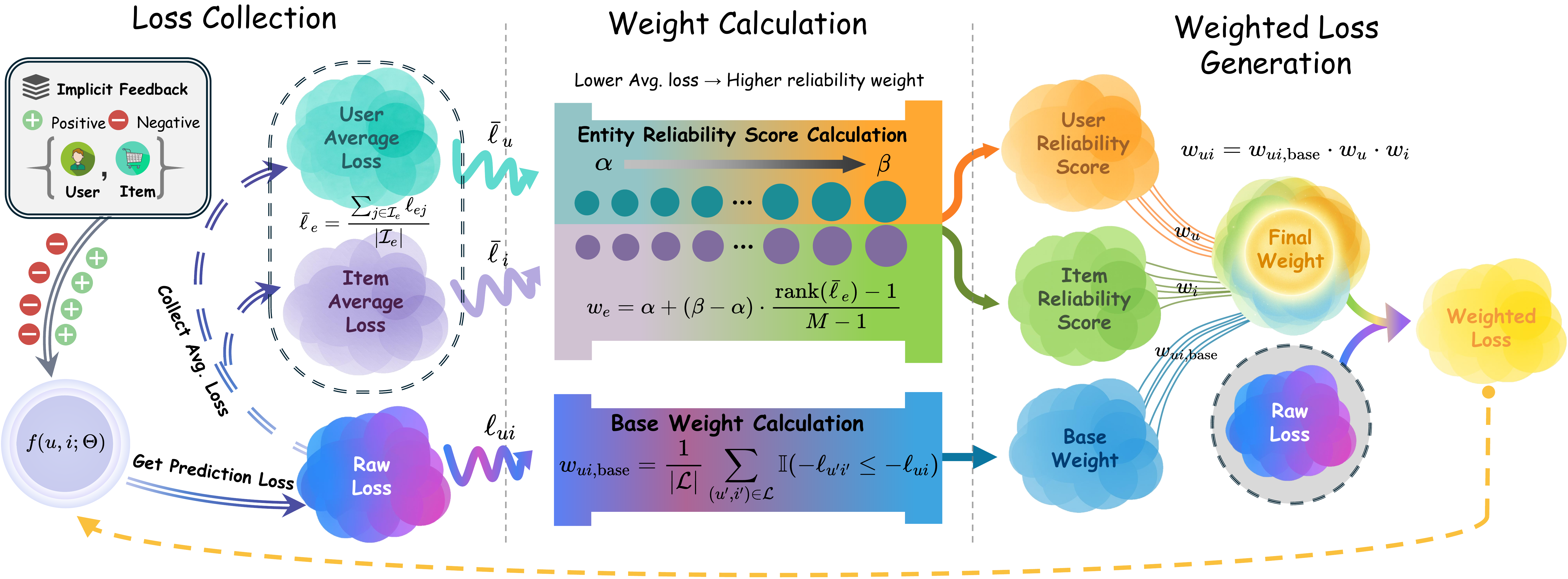}
\Description{Overview diagram of the EARD framework. It shows the three steps at the end of each epoch: Interaction Confidence, Entity Reliability, and Weight Fusion.}
\caption{\textbf{Overview of EARD.} At the end of each epoch, the framework performs three steps: (1) \textit{Collect Loss}: gather losses for all interactions and compute average loss per entity; (2) \textit{Entity Reliability and Base Weight Calculation}: estimate entity reliability via a linear mapping of average loss from $\alpha$ to $\beta$, and assign base weights using the ECDF of negative losses; (3) \textit{Weight Fusion}: combine interaction and entity weights through element-wise multiplication to generate training weights for the next epoch.}

\label{fig:whole_pipeline}
\end{figure*}

\subsubsection{Denoising as a Weighted Optimization Problem}

Recognizing the prevalence of noise, we formulate denoising as a weighted optimization problem. We associate a confidence weight $w_{ui} \in \mathbb{R}$ with each training instance $(u,i) \in \mathcal{D}$. The objective is to learn the model parameters $\Theta$ by minimizing the weighted loss over the set of training instances:

\begin{equation}
\label{eq:objective}
\min_{\Theta} \sum_{(u,i) \in \mathcal{D}} w_{ui} \cdot \ell(f(u, i; \Theta), y_{ui})
\end{equation}

\subsection{Core Assumptions}

Our method relies on two key observations: \textbf{1)} Noisy interactions tend to produce higher training losses, as supported by prior work in curriculum and robust learning~\cite{Co-teaching, MentorNet, Curriculumlearning} and illustrated in Figure~\ref{fig:full_TP_FP_loss}; \textbf{2)} Interaction reliability varies significantly across entities, as shown in Figure~\ref{fig:ml_1m_noisy_rate_distribution} and Appendix Figure~\ref{fig:all_noisy_rate_distribution}. These insights motivate our entity-aware design, which weights interactions based on reliability estimated from average training loss.

\subsection{The EARD Framework}

Building on these assumptions, \textbf{EARD} assigns weights to user, item, and interaction factors based on their estimated noise levels, and integrates them into a unified final weight.

\subsubsection{Factor-Driven Weighting Mechanism}

As illustrated in Figure~\ref{fig:whole_pipeline}, \textbf{EARD} operates iteratively. At the end of each epoch $t$, it computes a weight $w_{ui}$ for every instance $(u,i)$ in the training set $\mathcal{D}^{(t)}$. The design is stateless and computationally efficient. For clarity, we use italic lowercase for scalar weights ($w_{ui}$), and bold lowercase for entity-level reliability score vectors ($\mathbf{w}_u, \mathbf{w}_i$).

\paragraph{Step 1: Base Weight Generation ($w_{ui, \text{base}}$).}

For each training instance $(u,i) \in \mathcal{D}^{(t)}$, we compute a base weight $w_{ui, \text{base}}$ to quantify its interaction-level factor. This is obtained by applying the Empirical Cumulative Distribution Function (ECDF) to the loss set, a non-parametric method that is robust to outliers. The base weight for instance $(u,i)$ is defined as:

\begin{equation}
\label{eq:base_weight}
w_{ui, \text{base}}^{(t+1)} = \frac{1}{|\mathcal{L}^{(t)}|} \sum_{(u', i') \in \mathcal{L}^{(t)}} \mathbb{I}(-\ell_{u'i'}^{(t)} \le -\ell_{ui}^{(t)}) ,
\end{equation}

where $\mathcal{L}^{(t)}$ is the collection of all losses from the training set $\mathcal{D}^{(t)}$ in epoch $t$. This process assigns higher weights to lower-loss samples without storing a dense matrix.

\paragraph{Step 2: Entity-Aware Weight Computation \textnormal{($\mathbf{w}_u, \mathbf{w}_i$)}.}

We estimate user and item reliability using their average training loss in epoch $t$. For user $u$, the average loss is
\begin{equation}
\bar{\ell}_u^{(t)} = \frac{\sum_{i \in \mathcal{I}_u^{(t)}} \ell_{ui}^{(t)}}{|\mathcal{I}_u^{(t)}|},
\end{equation}
where $\mathcal{I}_u^{(t)}$ is the set of items interacted with by user $u$ during epoch $t$. The same applies to items. Assuming lower average loss indicates higher reliability, we map these values to reputation scores using a rank-based linear function. Let $\text{rank}(\bar{\ell}_u^{(t)})$ denote the ascending rank of user $u$'s average loss. Then the user reputation factor is
\begin{equation}
\label{eq:user_fact}
w_u^{(t+1)} = \alpha + (\beta - \alpha) \cdot \frac{\text{rank}(\bar{\ell}_u^{(t)}) - 1}{M - 1},
\end{equation}
where $\alpha, \beta \in [0, \infty)$ are hyperparameters and $\alpha \le \beta$. This mapping assigns the highest score $\beta$ to the most reliable entity (lowest loss) and $\alpha$ to the least. We repeat this process for all users and items to obtain reliability score vectors $\mathbf{w}_u^{(t+1)} \in \mathbb{R}^M$ and $\mathbf{w}_i^{(t+1)} \in \mathbb{R}^N$, which are then used for the final weight computation.

\paragraph{Step 3: Final Weight Fusion \textnormal{($w_{ui}$)}.}
We obtain the final weight $w_{ui}^{(t+1)}$ for each training instance $(u,i) \in \mathcal{D}^{(t+1)}$ by fusing its interaction-level confidence and the reliability of its associated entities. This is performed via a simple scalar multiplication:
\begin{equation}
\label{eq:final_weight}
w_{ui}^{(t+1)} = w_{ui, \text{base}}^{(t+1)} \cdot w_u^{(t+1)} \cdot w_i^{(t+1)}
\end{equation}
This formulation ensures high confidence only when both the interaction and its associated entities are considered reliable.

\subsection{Complexity Analysis}

We analyze the computational complexity of EARD to demonstrate its efficiency and scalability. Let $|\mathcal{D}| = |\mathcal{O}| + |\mathcal{S}|$ denote the total number of training samples per epoch.

\paragraph{Time Complexity.} EARD introduces three additional operations per epoch: (1) computing average losses in $O(|\mathcal{D}|)$; (2) generating reliability scores by sorting, costing $O(M \log M + N \log N)$; and (3) computing ECDF-based weights, requiring $O(|\mathcal{D}| \log |\mathcal{D}|)$.

\paragraph{Space Complexity.} The method maintains the loss and weight for all samples (i.e., valid interactions) within each training epoch, leading to a space complexity of $O(|\mathcal{D}|)$.

\section{Experiments}

\subsection{Experimental Setup}

We empirically evaluate the effectiveness and efficiency of \textbf{EARD} through the following research questions:

\noindent \textbf{1) RQ1 (Overall Performance):} Does {EARD} consistently outperform SOTA denoising methods across models and datasets?

\noindent \textbf{3) RQ2 (Ablation Study):} How do the entity-level and interaction-level weights contribute to overall performance?

\noindent \textbf{2) RQ3 (Mechanism Insight):} Can {EARD} effectively distinguish true-positive (clean) and false-positive (noisy) interactions based on the learned interaction and entity weights?

\noindent \textbf{4) RQ4 (Hyperparameter Sensitivity):} How sensitive is EARD to its hyperparameters $\alpha$ and $\beta$, and is it easy to tune?

\noindent \textbf{5) RQ5 (Efficiency):} What is the computational overhead of EARD, and is it scalable to large datasets? We report peak memory usage and training epochs until convergence, comparing with vanilla training and other denoising methods.


\subsubsection{Datasets.}

We evaluate our framework on three public datasets: \textbf{Yelp}\footnote{\url{https://www.yelp.com/dataset}}, \textbf{Amazon-Book}\footnote{\url{https://jmcauley.ucsd.edu/data/amazon/}}, and \textbf{ML-1M}\footnote{\url{https://grouplens.org/datasets/movielens/1m/}}. Following~\cite{ADT}, we binarize interactions by treating ratings $\le 3$ as \textit{false positives} (noisy) and $>3$ as \textit{true positives}. \textbf{Yelp} contains user reviews for local businesses.  \textbf{Amazon-Book} consists of user ratings for books.  \textbf{ML-1M} includes movie ratings from the GroupLens research group.

\begin{table}[htbp]
\centering
\small
\setlength{\tabcolsep}{1mm}
\caption{
Dataset statistics. ``\#False Positives'' denotes noisy interactions (ratings $\le 3$) addressed by our framework.
}
\resizebox{\linewidth}{!}{%
\begin{tabular}{@{}lcccccc@{}}
\toprule
\textbf{Dataset} & \textbf{\#Users} & \textbf{\#Items} & \textbf{\#Interactions} & \textbf{\#False Positives} & \textbf{Sparsity} \\
\midrule
Yelp & 45,548 & 57,396 & 1,672,520 & 260,581 & 0.0640\% \\
Amazon-book & 80,464 & 98,663 & 2,714,021 & 199,475 & 0.0342\% \\
ML-1M & 6,040 & 3,953 & 706,391 & 338,746 & 2.959\% \\
\bottomrule
\end{tabular}
}
\label{tab:dataset_statistics}
\end{table}

All datasets are randomly split into training, validation, and test sets in an \textbf{8:1:1} ratio. Key statistics are summarized in Table~\ref{tab:dataset_statistics}.

\subsubsection{Backbone Models.}
To demonstrate the model-agnostic nature of EARD, we apply it to three representative collaborative filtering models: \textbf{GMF}~\cite{NeuMF}, which uses element-wise multiplication of user and item embeddings; \textbf{NeuMF}~\cite{NeuMF}, which combines GMF with multilayer perceptrons to model both linear and non-linear interactions; and \textbf{CDAE}~\cite{CDAE}, an autoencoder that reconstructs user interaction vectors from corrupted inputs.

\begin{table*}[!t]
\centering
\small
\caption{Overall performance of three representative backbone trained using seven awesome different denoising methods. `Recall@K' and `NDCG@K' are shortened as `R@K' and `N@K'. `$\Delta$\%' refers to the relative improvement of our method compared to the other baselines. `*' indicates statistically significant improvement by t-test (p $<$ 0.05) to other methods.}
\resizebox{\textwidth}{!}{
\begin{tabular}{@{}llcccccccccccc@{}}
\toprule
\multirow{2}{*}{\textbf{Model}} & \multirow{2}{*}{\textbf{Method}} & \multicolumn{4}{c}{\textbf{ML-1M}} & \multicolumn{4}{c}{\textbf{Yelp}} & \multicolumn{4}{c}{\textbf{Amazon-book}} \\
\cmidrule(lr){3-6} \cmidrule(lr){7-10} \cmidrule(lr){11-14}
& & \textbf{R@50} & \textbf{R@100} & \textbf{N@50} & \textbf{N@100} & \textbf{R@50} & \textbf{R@100} & \textbf{N@50} & \textbf{N@100} & \textbf{R@50} & \textbf{R@100} & \textbf{N@50} & \textbf{N@100} \\
\midrule
\multirow{8}{*}{GMF} & Base & 0.3967 & 0.5523 & 0.3375 & 0.3970 & 0.1020 & 0.1673 & 0.0406 & 0.0552 & 0.0851 & 0.1347 & 0.0341 & 0.0450 \\
& WRMF & 0.3785 & 0.5278 & 0.3286 & 0.3849 & 0.0875 & 0.1427 & 0.0363 & 0.0487 & 0.0734 & 0.1157 & 0.0301 & 0.0394 \\
& R-CE & 0.3749 & 0.5240 & 0.3277 & 0.3840 & 0.0864 & 0.1365 & 0.0367 & 0.0481 & 0.0657 & 0.1041 & 0.0268 & 0.0353 \\
& T-CE & 0.3686 & 0.5226 & 0.3154 & 0.3739 & 0.0900 & 0.1471 & 0.0363 & 0.0493 & 0.0738 & 0.1147 & 0.0306 & 0.0397 \\
& DeCA & 0.2757 & 0.4007 & 0.2294 & 0.2778 & 0.0466 & 0.0823 & 0.0169 & 0.0248 & 0.0616 & 0.1050 & 0.0220 & 0.0314 \\
& BOD & 0.4030 & 0.5569 & {0.3427} & {0.4014} & 0.0954 & 0.1559 & 0.0387 & 0.0522 & 0.0762 & 0.1223 & 0.0302 & 0.0404 \\
& PLD & 0.4121 & \underline{0.5716} & \underline{0.3517} & \underline{0.4128} & 0.0982 & 0.1610 & 0.0394 & 0.0535 & 0.0779 & 0.1251 & 0.0310 & 0.0414 \\
& UDT & \underline{0.4131} & {0.5585} & 0.3426 & 0.3987 & \underline{0.1352} & \underline{0.2118} & \underline{0.0569} & \underline{0.0741} & \textbf{0.1228} & \textbf{0.1807} & \textbf{0.0524} & \textbf{0.0653} \\
& Ours & \textbf{0.4342*} & \textbf{0.5851*} & \textbf{0.3667*} & \textbf{0.4241*} & \textbf{0.1375*} & \textbf{0.2169*} & \textbf{0.0575*} & \textbf{0.0753*} & \underline{0.1193} & \underline{0.1765} & \underline{0.0506} & \underline{0.0633} \\
\cmidrule(l){2-14}
& $\Delta$\% & 5.11\% & 2.36\% & 4.26\% & 2.74\% & 1.70\% & 2.41\% & 1.05\% & 1.62\% & -2.85\% & -2.32\% & -3.44\% & -3.06\% \\
\midrule
\multirow{8}{*}{NeuMF} & Base & 0.3868 & 0.5431 & 0.3171 & 0.3770 & 0.0875 & 0.1516 & 0.0332 & 0.0475 & 0.0844 & 0.1341 & 0.0330 & 0.0439 \\
& WRMF & 0.3959 & 0.5495 & 0.3300 & 0.3889 & 0.0806 & 0.1353 & 0.0318 & 0.0441 & 0.0711 & 0.1141 & 0.0280 & 0.0374 \\
& R-CE & 0.3803 & 0.5367 & 0.3188 & 0.3782 & 0.0803 & 0.1306 & 0.0325 & 0.0438 & 0.0606 & 0.0970 & 0.0244 & 0.0324 \\
& T-CE & 0.3943 & 0.5533 & 0.3205 & 0.3820 & 0.0831 & 0.1391 & 0.0321 & 0.0447 & 0.0693 & 0.1106 & 0.0274 & 0.0365 \\
& DeCA & 0.2420 & 0.3510 & 0.2109 & 0.2516 & 0.0464 & 0.0904 & 0.0162 & 0.0258 & 0.0420 & 0.0792 & 0.0144 & 0.0222 \\
& BOD & 0.3782 & 0.5242 & 0.3262 & 0.3816 & 0.0851 & 0.1371 & 0.0349 & 0.0467 & 0.0768 & 0.1262 & 0.0287 & 0.0396 \\
& PLD & 0.3847 & 0.5339 & \underline{0.3345} & \underline{0.3920} & 0.0866 & 0.1398 & 0.0357 & 0.0482 & 0.0791 & 0.1293 & 0.0295 & 0.0406 \\
& UDT & \underline{0.4126} & \underline{0.5649} & {0.3314} & {0.3901} & \underline{0.1085} & \underline{0.1790} & \underline{0.0422} & \underline{0.0580} & \underline{0.1091} & \underline{0.1650} & \underline{0.0459} & \underline{0.0583} \\
& Ours & \textbf{0.4257*} & \textbf{0.5747*} & \textbf{0.3617*} & \textbf{0.4181*} & \textbf{0.1306*} & \textbf{0.2057*} & \textbf{0.0536*} & \textbf{0.0704*} & \textbf{0.1235*} & \textbf{0.1820*} & \textbf{0.0530*} & \textbf{0.0658*} \\
\cmidrule(l){2-14}
& $\Delta$\% & 3.17\% & 1.73\% & 8.13\% & 6.66\% & 20.37\% & 14.92\% & 27.01\% & 21.38\% & 13.20\% & 10.30\% & 15.47\% & 12.86\% \\
\midrule
\multirow{8}{*}{CDAE} & Base & 0.4299 & 0.5802 & 0.3656 & 0.4252 & 0.1165 & 0.1877 & 0.0466 & 0.0625 & 0.1024 & 0.1574 & 0.0417 & 0.0538 \\
& WRMF & 0.4095 & 0.5590 & 0.3505 & 0.4081 & 0.0943 & 0.1534 & 0.0375 & 0.0507 & 0.0896 & 0.1372 & 0.0377 & 0.0438 \\
& R-CE & 0.4114 & 0.5600 & 0.3565 & 0.4135 & 0.1161 & 0.1801 & 0.0488 & 0.0632 & 0.1022 & 0.1560 & 0.0424 & 0.0542 \\
& T-CE & 0.4033 & 0.5572 & 0.3394 & 0.3994 & 0.1165 & 0.1806 & 0.0504 & 0.0652 & 0.1088 & 0.1645 & 0.0454 & 0.0575 \\
& DeCA & 0.2446 & 0.3293 & 0.1763 & 0.2123 & 0.0767 & 0.1212 & 0.0306 & 0.0409 & 0.0663 & 0.1040 & 0.0265 & 0.0343 \\
& BOD & 0.4161 & 0.5668 & 0.3542 & 0.4125 & 0.1115 & 0.1755 & 0.0457 & 0.0601 & 0.0911 & 0.1409 & 0.0374 & 0.0482 \\
& PLD & 0.4296 & \underline{0.5836} & 0.3620 & 0.4200 & 0.1144 & 0.1804 & 0.0468 & 0.0618 & 0.0932 & 0.1441 & 0.0383 & 0.0492 \\
& UDT & \underline{0.4347} & {0.5818} & \underline{0.3749} & \underline{0.4317} & \underline{0.1321} & \underline{0.2013} & \underline{0.0566} & \underline{0.0722} & \underline{0.1219} & \underline{0.1768} & \underline{0.0525} & \underline{0.0647} \\
& Ours & \textbf{0.4500*} & \textbf{0.5930*} & \textbf{0.3900*} & \textbf{0.4447*} & \textbf{0.1562*} & \textbf{0.2346*} & \textbf{0.0675*} & \textbf{0.0850*} & \textbf{0.1447*} & \textbf{0.2073*} & \textbf{0.0633*} & \textbf{0.0770*} \\
\cmidrule(l){2-14}
& $\Delta$\% & 3.52\% & 1.61\% & 4.03\% & 3.01\% & 18.24\% & 16.54\% & 19.26\% & 17.73\% & 18.70\% & 17.25\% & 20.57\% & 19.01\% \\

\bottomrule
\end{tabular}
}
\label{tab:overall_performance}
\end{table*}

\subsubsection{Evaluation Protocol.}

We follow standard practice by ranking all non-interacted items for each user. Evaluation focuses on \textit{true-positive} test set, reflecting the goal of recommending items users genuinely prefer. We report two common top-K ranking metrics: \textbf{Recall@K} and \textbf{NDCG@K}, with K=\{50, 100\}.

\subsubsection{Baselines.}

We compared EARD against backbone models and representative denoising baselines, including WRMF~\cite{WRMF}, R-CE~\cite{ADT}, T-CE~\cite{ADT}, model-disagreement approaches like DeCA~\cite{DeCA}, BOD~\cite{BOD}, UDT~\cite{UDT} and PLD~\cite{PLD}. \footnote{We do not compare with LLaRD~\cite{LLaRD}, as it relies on additional side information such as item descriptions, user reviews, and external world knowledge from large language models. These resources are unavailable to other baselines and inconsistent with our experimental setup focused on interaction-only data. We also exclude SGDL~\cite{SGDL} from comparison, as it fails to complete training within a reasonable time on datasets with over one million interactions.}

\subsubsection{Implementation Details}

All experiments were conducted on an NVIDIA 4090 GPU using the Adam optimizer~\cite{Adam}. The embedding size was set to 32, the learning rate to $10^{-3}$, the batch size to 2048, and the negative sampling rate ($k$) to 1. For EARD, we tuned the hyperparameters $\alpha$ and $\beta$ within the range $[0, 5]$. The optimal configurations are reported in Table~\ref{tab:optimal_hyperparameters} in Appendix~\ref{sec:hyperparameterAnalysis}. To ensure stability and reliability, all results are averaged over five independent runs.

\subsection{Performance Comparison (RQ1)}

We benchmarked \textbf{EARD} against state-of-the-art baselines across all datasets and backbone models, using their best reported configurations for a fair comparison.

As shown in Table~\ref{tab:overall_performance}, EARD consistently outperforms all baselines. Heuristic methods such as R-CE and T-CE yield unstable results and occasionally underperform the backbone, indicating that naively pruning high-loss samples can discard informative but challenging instances. More complex approaches like DeCA also underperform, suggesting that added architectural complexity does not guarantee improved denoising. In contrast, EARD achieves robust gains by leveraging entity-level attribution, with particularly strong improvements on larger datasets where performance gaps are more pronounced. Notably, even on CDAE, which is inherently robust to noise, EARD delivers significant improvements (e.g., +19.26\% and +20.57\% NDCG@50 on Yelp and Amazon-Book), demonstrating its complementarity to model-based robustness. 

Compared to UDT, EARD generally achieves superior performance, especially on advanced models like NeuMF and CDAE. The only exception is GMF on Amazon-Book, where UDT slightly outperforms it. On NeuMF, EARD delivers an average NDCG@50 improvement of up to \textbf{27.01\%} on Yelp. It also requires fewer hyperparameters (Ours: 2 vs. UDT: 4), making it more efficient to tune and better suited for practical deployment. These results highlight the advantage of entity-level modeling for denoising implicit feedback.

\subsection{Ablation Study (RQ2)}

\subsubsection{Component Contribution Analysis}

\begin{table*}[!t]
\centering
\small
\caption{Ablation results on ML-1M, Yelp, and Amazon-Book for GMF, NeuMF, and CDAE. We evaluate three components: ECDF-based Weight (BW), Item Weight (IF), and User Weight (UF). \checkmark indicates active, $\times$ indicates inactive.}
\resizebox{\textwidth}{!}{
\begin{tabular}{@{}cccccccccccccccc@{}}
\toprule
\multirow{2}{*}{\textbf{Model}} & \multicolumn{3}{c}{\textbf{Components}} & \multicolumn{4}{c}{\textbf{ML-1M}} & \multicolumn{4}{c}{\textbf{Yelp}} & \multicolumn{4}{c}{\textbf{Amazon-book}} \\
\cmidrule(lr){2-4} \cmidrule(lr){5-8} \cmidrule(lr){9-12} \cmidrule(lr){13-16}
& \textbf{BW} & \textbf{IF} & \textbf{UF} & \textbf{R@50} & \textbf{R@100} & \textbf{N@50} & \textbf{N@100} & \textbf{R@50} & \textbf{R@100} & \textbf{N@50} & \textbf{N@100} & \textbf{R@50} & \textbf{R@100} & \textbf{N@50} & \textbf{N@100} \\
\midrule
\multirow{5}{*}{GMF} 
& $\times$ & $\times$ & $\times$ & 0.3967 & 0.5523 & 0.3375 & 0.3970 & 0.1020 & 0.1673 & 0.0406 & 0.0552 & 0.0851 & 0.1347 & 0.0341 & 0.0450 \\
& \checkmark & $\times$ & $\times$ & 0.4271 & 0.5765 & 0.3593 & 0.4166 & 0.1373 & 0.2158 & 0.0575 & 0.0750 & 0.1143 & 0.1745 & 0.0471 & 0.0604 \\
& \checkmark & \checkmark & $\times$ & \textbf{0.4353} & \underline{0.5821} & \textbf{0.3681} & \textbf{0.4242} & 0.1373 & \underline{0.2164} & 0.0574 & 0.0751 & 0.1190 & \underline{0.1795} & 0.0498 & 0.0631 \\
& \checkmark & $\times$ & \checkmark & 0.4286 & 0.5813 & 0.3636 & 0.4218 & \textbf{0.1380} & 0.2163 & \textbf{0.0577} & \underline{0.0752} & \textbf{0.1195} & \textbf{0.1806} & \underline{0.0499} & \underline{0.0633} \\
& \checkmark & \checkmark & \checkmark & \underline{0.4342} & \textbf{0.5851} & \underline{0.3667} & \underline{0.4241} & \underline{0.1375} & \textbf{0.2169} & \underline{0.0575} & \textbf{0.0753} & \underline{0.1193} & 0.1765 & \textbf{0.0506} & \textbf{0.0633} \\
\midrule
\multirow{5}{*}{NeuMF} 
& $\times$ & $\times$ & $\times$ & 0.3868 & 0.5431 & 0.3171 & 0.3770 & 0.0875 & 0.1516 & 0.0332 & 0.0475 & 0.0844 & 0.1341 & 0.0330 & 0.0439 \\
& \checkmark & $\times$ & $\times$ & 0.4226 & 0.5740 & 0.3605 & 0.4176 & 0.1040 & 0.1713 & 0.0410 & 0.0516 & 0.0912 & 0.1445 & 0.0361 & 0.0479 \\
& \checkmark & \checkmark & $\times$ & 0.4252 & 0.5739 & \textbf{0.3623} & \textbf{0.4184} & \underline{0.1238} & \underline{0.1984} & \underline{0.0494} & \underline{0.0660} & 0.1104 & 0.1677 & 0.0453 & 0.0578 \\
& \checkmark & $\times$ & \checkmark & 0.4225 & 0.5738 & 0.3595 & 0.4166 & 0.1203 & 0.1959 & 0.0480 & 0.0649 & 0.1086 & 0.1676 & 0.0446 & 0.0576 \\
& \checkmark & \checkmark & \checkmark & \textbf{0.4257} & \textbf{0.5747} & \underline{0.3617} & \underline{0.4181} & \textbf{0.1306} & \textbf{0.2057} & \textbf{0.0536} & \textbf{0.0704} & \textbf{0.1235} & \textbf{0.1820} & \textbf{0.0530} & \textbf{0.0658} \\
\midrule
\multirow{5}{*}{CDAE} 
& $\times$ & $\times$ & $\times$ & 0.4299 & 0.5802 & 0.3656 & 0.4252 & 0.1165 & 0.1877 & 0.0466 & 0.0625 & 0.1024 & 0.1574 & 0.0417 & 0.0538 \\
& \checkmark & $\times$ & $\times$ & 0.4390 & 0.5871 & 0.3794 & 0.4365 & 0.1509 & 0.2299 & 0.0636 & 0.0812 & 0.1389 & 0.2030 & 0.0592 & 0.0731 \\
& \checkmark & \checkmark & $\times$ & 0.4405 & 0.5839 & \underline{0.3855} & \underline{0.4404} & 0.1535 & 0.2318 & 0.0652 & 0.0827 & 0.1426 & 0.2055 & 0.0619 & 0.0756 \\
& \checkmark & $\times$ & \checkmark & \underline{0.4447} & \underline{0.5918} & 0.3820 & 0.4390 & \textbf{0.1573} & \textbf{0.2371} & \underline{0.0664} & \underline{0.0841} & \textbf{0.1458} & \textbf{0.2101} & \underline{0.0633} & \textbf{0.0773} \\
& \checkmark & \checkmark & \checkmark & \textbf{0.4500} & \textbf{0.5930} & \textbf{0.3900} & \textbf{0.4447} & \underline{0.1562} & \underline{0.2346} & \textbf{0.0675} & \textbf{0.0850} & \underline{0.1447} & \underline{0.2073} & \textbf{0.0633} & \underline{0.0770} \\
\bottomrule
\end{tabular}
}
\label{tab:components_ablation_results_full}
\end{table*}

We conducted an ablation study (Table~\ref{tab:components_ablation_results_full}) to evaluate the contribution of each component in EARD, which comprises three elements: the interaction-specific Base Weight (\textbf{BW}), Item Factor (\textbf{IF}), and User Factor (\textbf{UF}). Each component was selectively disabled, and performance was measured using Recall@50/100 and NDCG@50/100.

\paragraph{The Foundational Role of Interaction-level Weighting (BW).}
Enabling only BW (\texttt{\checkmark},$\times$,$\times$) yields a clear and substantial improvement over the vanilla baseline ($\times$,$\times$,$\times$) across all models and datasets. For instance, on NeuMF with the ML-1M dataset, relying solely on BW increases NDCG@50 from 0.3171 to 0.3605, a relative gain of \textbf{13.7\%}. This result strongly substantiates our core hypothesis: an interaction’s training loss serves as a powerful and reliable proxy for its underlying quality and can be directly exploited for effective denoising, forming the foundation of our framework.

\paragraph{Complementary and Context-Dependent Effects of Entity Factors (IF \& UF).}
Introducing either the Item Factor or the User Factor on top of BW almost always brings further performance gains, demonstrating the critical value of entity-aware modeling. The effects of IF and UF appear to be complementary and context-dependent. For example, the \textbf{User Factor} (\texttt{\checkmark},$\times$,\texttt{\checkmark}) shows a particularly strong impact on the CDAE model with the Amazon-Book dataset, suggesting that when user behaviors are highly diverse and noisy, explicitly modeling user-specific reliability is crucial. Conversely, the \textbf{Item Factor} (\texttt{\checkmark},\texttt{\checkmark},$\times$) achieves the best performance on three out of four metrics for GMF on ML-1M, indicating that in scenarios where item-side noise (e.g., clickbait or low-quality items) is dominant, identifying unreliable items is a primary driver of improvement.


\paragraph{Synergy in Full Configuration and Its Implications.}
The full EARD configuration (\texttt{\checkmark},\texttt{\checkmark},\texttt{\checkmark}) generally delivers top-tier performance, highlighting the strong synergy among its components. BW provides a solid foundation for interaction quality, while IF and UF contribute personalized and orthogonal signals that enhance reliability. For instance, on the Yelp dataset with NeuMF, the full model achieves an NDCG@50 of 0.0536, outperforming BW-only by 30.7\%, BW+IF by 8.5\%, and BW+UF by 11.7\%. These results demonstrate that modeling all three factors together enables EARD to generalize effectively in sparse and noisy environments.

\subsubsection{Exploration of Base Weight Calculation Methods}

\begin{table*}[!t]
\centering
\small
\caption{Full evaluation of different base weighting strategies (Uniform, GMM, Top-$k$ Selection, Linear Scaling, ECDF) integrated into the EARD framework, across three datasets and three models. Metrics include Recall@50/100 and NDCG@50/100. ECDF consistently achieves the highest performance, validating its effectiveness and robustness.}
\resizebox{\textwidth}{!}{
\begin{tabular}{llcccccccccccc}
\toprule
\multirow{2}{*}{\textbf{Model}} & \multirow{2}{*}{\textbf{Method}} & \multicolumn{4}{c}{\textbf{ML-1M}} & \multicolumn{4}{c}{\textbf{Yelp}} & \multicolumn{4}{c}{\textbf{Amazon-Book}} \\
\cmidrule(lr){3-6} \cmidrule(lr){7-10} \cmidrule(lr){11-14}
& & \textbf{R@50} & \textbf{R@100} & \textbf{N@50} & \textbf{N@100} & \textbf{R@50} & \textbf{R@100} & \textbf{N@50} & \textbf{N@100} & \textbf{R@50} & \textbf{R@100} & \textbf{N@50} & \textbf{N@100} \\
\midrule
\multirow{5}{*}{GMF} & Uniform & 0.3844 & 0.5393 & 0.3219 & 0.3815 & 0.1028 & 0.1679 & 0.0410 & 0.0555 & 0.1136 & 0.1743 & 0.0474 & 0.0608 \\
& Top-$k$ Selection & 0.4011 & 0.5668 & 0.3161 & 0.3806 & 0.1266 & 0.1995 & 0.0530 & 0.0695 & 0.0980 & 0.1486 & 0.0415 & 0.0529 \\
& Linear Scaling & 0.3873 & 0.5421 & 0.3244 & 0.3889 & 0.1026 & 0.1682 & 0.0409 & 0.0556 & 0.1138 & 0.1742 & 0.0475 & 0.0608 \\
& GMM & 0.3215 & 0.4501 & 0.2853 & 0.3337 & 0.0879 & 0.1389 & 0.0367 & 0.0481 & 0.0389 & 0.0629 & 0.0157 & 0.0210 \\
& ECDF (Ours) & \textbf{0.4342} & \textbf{0.5851} & \textbf{0.3667} & \textbf{0.4241} & \textbf{0.1375} & \textbf{0.2169} & \textbf{0.0575} & \textbf{0.0753} & \textbf{0.1193} & \textbf{0.1765} & \textbf{0.0506} & \textbf{0.0633} \\
\midrule
\multirow{5}{*}{NeuMF} & Uniform & 0.4005 & 0.5510 & 0.3374 & 0.3944 & 0.1173 & 0.1874 & 0.0470 & 0.0626 & 0.1059 & 0.1604 & 0.0440 & 0.0559 \\
& Top-$k$ Selection & 0.4073 & 0.5629 & 0.3273 & 0.3868 & 0.1230 & 0.1930 & 0.0511 & 0.0669 & 0.1064 & 0.1567 & 0.0460 & 0.0573 \\
& Linear Scaling & 0.4023 & 0.5541 & 0.3396 & 0.3971 & 0.1190 & 0.1914 & 0.0482 & 0.0644 & 0.1103 & 0.1682 & 0.0459 & 0.0585 \\
& GMM & 0.3080 & 0.4364 & 0.2669 & 0.3164 & 0.0427 & 0.0701 & 0.0179 & 0.0241 & 0.0374 & 0.0612 & 0.0159 & 0.0213 \\
& ECDF (Ours) & \textbf{0.4257} & \textbf{0.5747} & \textbf{0.3617} & \textbf{0.4181} & \textbf{0.1306} & \textbf{0.2057} & \textbf{0.0536} & \textbf{0.0704} & \textbf{0.1235} & \textbf{0.1820} & \textbf{0.0530} & \textbf{0.0658} \\
\midrule
\multirow{5}{*}{CDAE} & Uniform & 0.4299 & 0.5779 & 0.3719 & 0.4292 & 0.1426 & 0.2190 & 0.0611 & 0.0781 & 0.1256 & 0.1853 & 0.0537 & 0.0667 \\
& Top-$k$ Selection & 0.4349 & 0.5780 & 0.3682 & 0.4238 & 0.1470 & 0.2213 & 0.0633 & 0.0800 & 0.1290 & 0.1849 & 0.0572 & 0.0696 \\
& Linear Scaling & 0.4323 & 0.5805 & 0.3736 & 0.4309 & 0.1430 & 0.2192 & 0.0613 & 0.0783 & 0.1249 & 0.1864 & 0.0533 & 0.0667 \\
& GMM & 0.3319 & 0.4684 & 0.2836 & 0.3371 & 0.1239 & 0.1864 & 0.0541 & 0.0682 & 0.1178 & 0.1687 & 0.0518 & 0.0629 \\
& ECDF & \textbf{0.4500} & \textbf{0.5930} & \textbf{0.3900} & \textbf{0.4447} & \textbf{0.1562} & \textbf{0.2346} & \textbf{0.0675} & \textbf{0.0850} & \textbf{0.1447} & \textbf{0.2073} & \textbf{0.0633} & \textbf{0.0770} \\
\bottomrule
\end{tabular}%
}
\label{tab:base_weight_method_comparison_full}
\end{table*}

\begin{figure}[!t]
\centering
\includegraphics[width=1.0\linewidth]{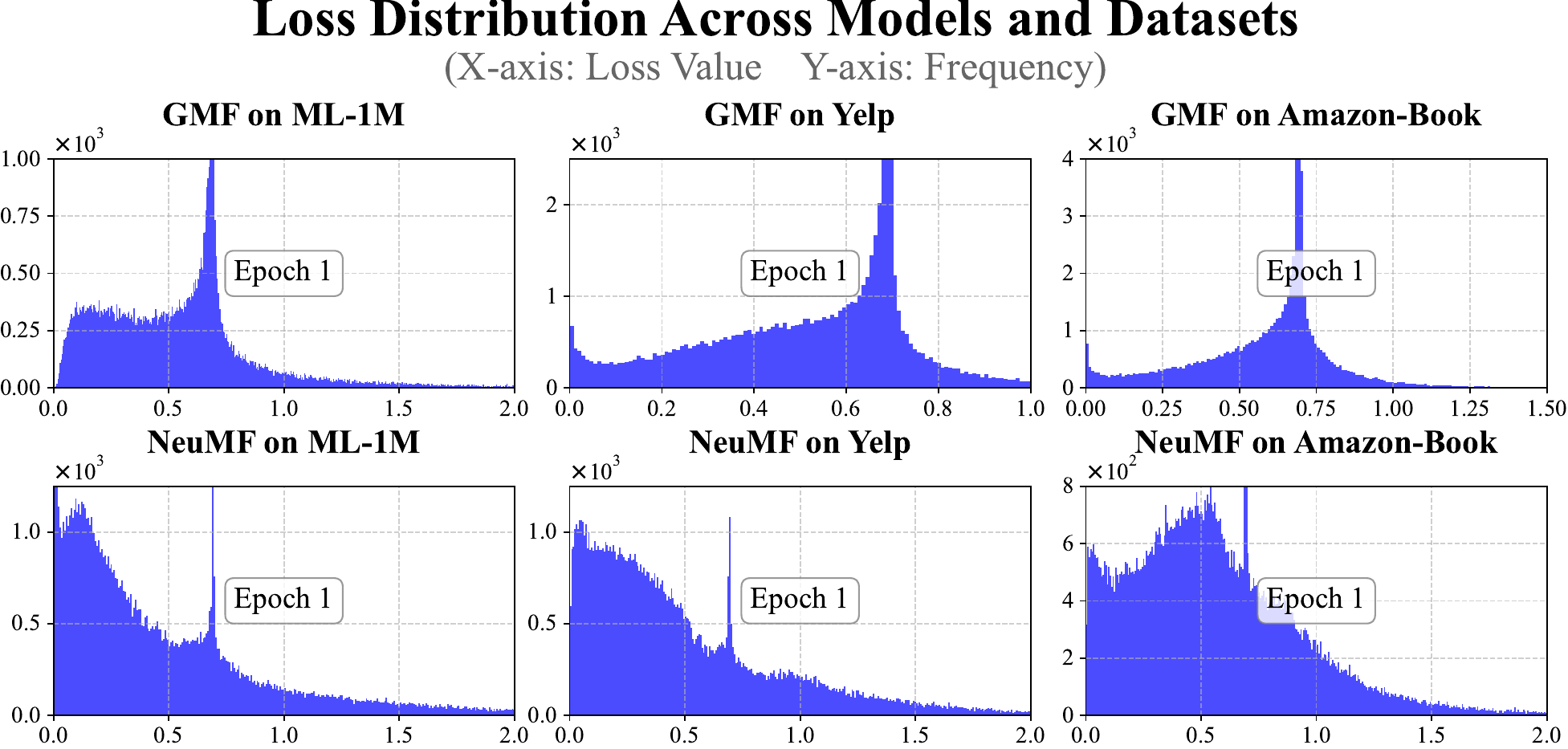}
\Description{Histogram showing training loss distributions at the end of the first epoch across different models and datasets.}
\caption{Training loss distributions at the end of the first epoch across different models and datasets.}
\label{fig:NeuMF_on_movielens_loss_distribution}
\end{figure}

The interaction-specific \textbf{Base Weight (BW)} plays a critical role in EARD. As shown in Figure~\ref{fig:NeuMF_on_movielens_loss_distribution} shows that the loss distributions of different models and datasets do not follow the same pattern and often contain long-tail outliers. This highlights the need for a weight calculation method that can adapt to diverse distribution shapes with a nonparametric weight function while enhancing robustness by reducing the influence of extreme values. ECDF is distribution-agnostic and resistant to the influence of outliers, since its output depends on relative rank rather than absolute magnitudes. Moreover, it provides smooth weights over the $[0,1]$ interval, avoiding the information loss caused by hard thresholds and yielding a stable denoising process. 

To validate its effectiveness, we compared ECDF against four alternative strategies: (1) \textbf{Uniform}: assigns a constant weight of 1 to all samples; (2) \textbf{GMM}: fits a Gaussian Mixture Model to the loss distribution; (3) \textbf{Top-$k$ Selection}: retains only the k lowest-loss samples per batch; (4) \textbf{Linear Scaling}: applies a direct linear transformation to the losses. Each strategy replaces only the ECDF component in the EARD framework. Pseudocode for all methods is provided in Appendix~\ref{sec:base_weight_computation_methods}.

As shown in Table~\ref{tab:base_weight_method_comparison_full}, ECDF consistently outperforms all alternative weighting strategies, confirming its effectiveness and robustness. The GMM-based method performs poorly due to its reliance on Gaussian assumptions, which fail under the diverse and non-standard loss distributions. Top-$k$ Selection applies a hard threshold, discarding high-loss samples entirely, which can eliminate informative but difficult interactions and leads to unstable training behavior. Linear Scaling directly maps loss magnitudes to weights, making it highly sensitive to outliers; a few large losses can dominate the weight range and distort the learning dynamics. In contrast, ECDF relies on loss rank rather than magnitude, making it robust to outliers and distribution-agnostic. Its smooth weighting over $[0,1]$ preserves fine-grained differences without abrupt cutoffs, making it well-suited for denoising implicit feedback.

\subsection{Denoising Mechanism Analysis (RQ3)}

\begin{figure}[!t]
\centering
\begin{subfigure}[t]{1.0\linewidth}
\centering
\includegraphics[width=1.0\linewidth]{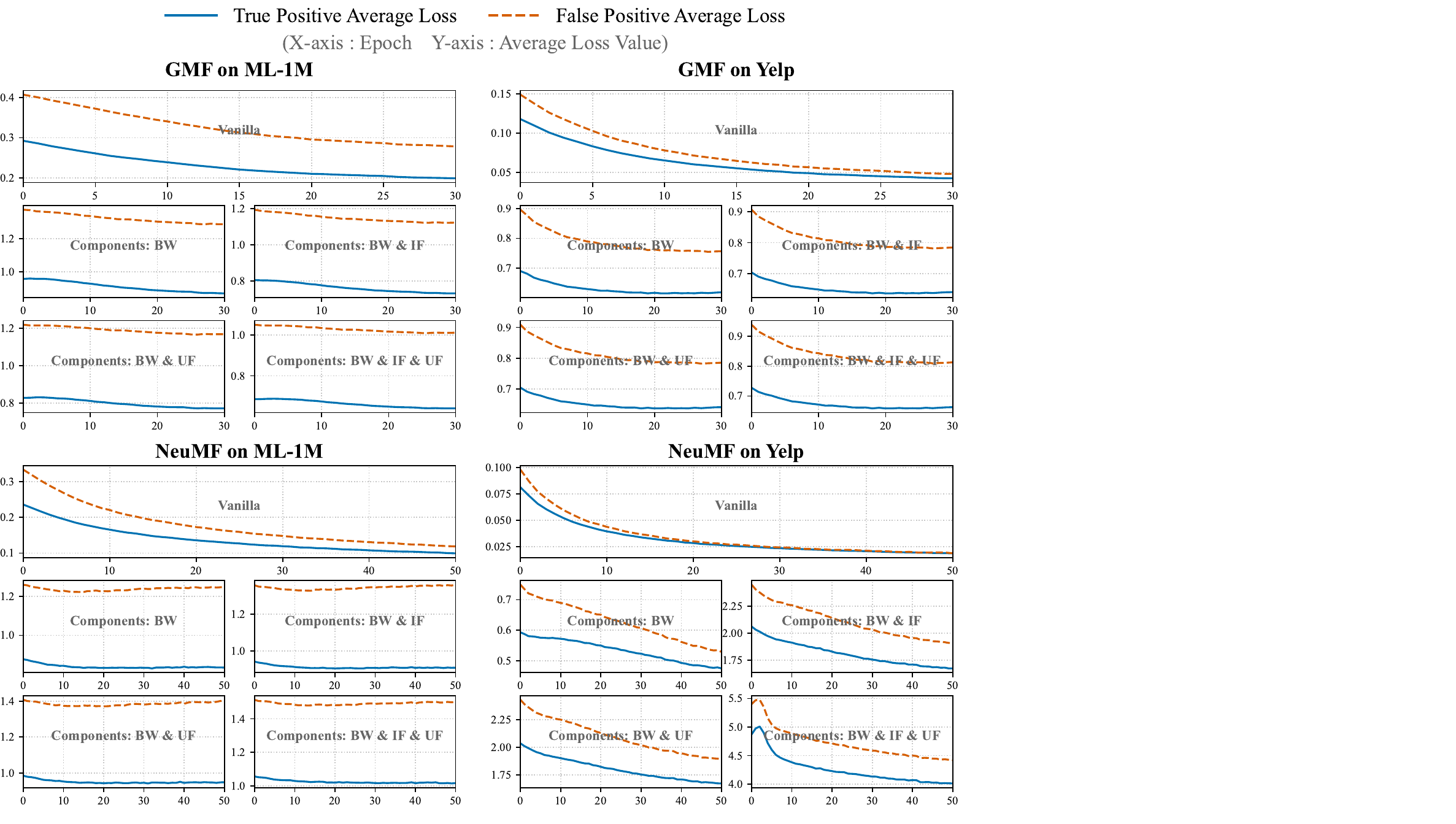}
\Description{Line plot showing epoch-wise average loss of True Positive and False Positive samples across different model settings on multiple datasets.}
\caption{}
\label{fig:full_TP_FP_loss}
\end{subfigure}
\begin{subfigure}[t]{1.0\linewidth}
\centering
\includegraphics[width=1.0\linewidth]{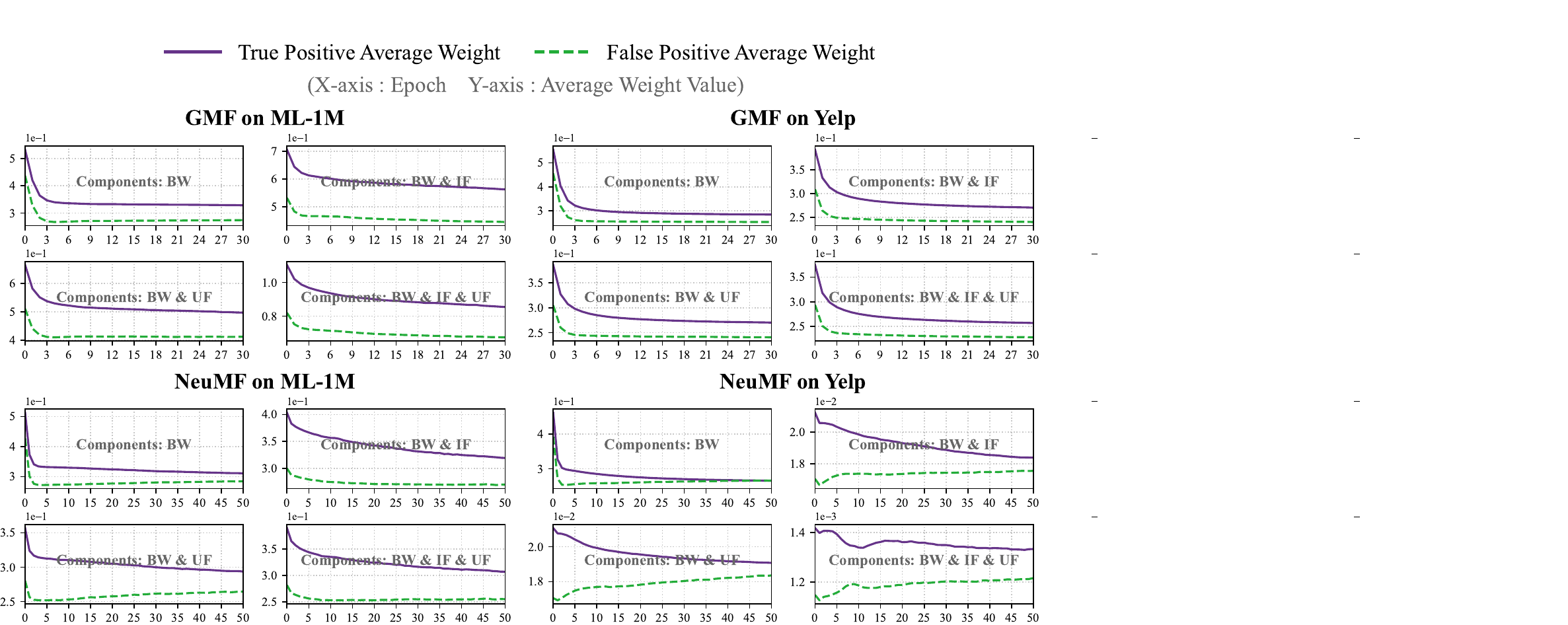}
\Description{Line plot showing average weight assignments for TP and FP samples during training across different model configurations.}
\caption{}
\label{fig:full_TP_FP_weight}
\end{subfigure}
\vspace{-1.0em}

\caption{Training dynamics of true positive (TP) and false positive (FP) samples across models and datasets. Base Weight (BW) is interaction-specific, while User Factor (UF) and Item Factor (IF) capture user and item reliability, respectively. (a) TP and FP loss over training epochs. EARD preserves a clear loss gap, with FP samples generally showing higher losses, supporting the small-loss assumption. (b) Assigned weights for TP and FP samples. EARD steadily gives higher weights to TP samples, highlighting its ability to identify clean interactions and suppress noise.}

\label{fig:full_TP_FP_loss_and_weight}
\end{figure}

We analyzed training loss and assigned weights for two groups: true positives (TP), representing clean interactions, and false positives (FP), representing noisy ones. As shown in Figure~\ref{fig:full_TP_FP_loss}, FP samples generally exhibit higher average losses than TP samples, supporting the small-loss assumption and suggesting that noisy interactions are harder to fit. In contrast, the vanilla model eventually reduces both TP and FP losses toward zero, indicating that it overfits the noise in the absence of denoising.

EARD amplifies the separation between true positive (TP) and false positive (FP) samples during training. As shown in Figure~\ref{fig:full_TP_FP_weight}, models with our framework maintain a clear and stable gap in assigned weights between TP and FP samples across architectures (GMF and NeuMF) and datasets (ML-1M and Yelp). This separation holds across all component settings (e.g., BW, BW \& IF, BW \& UF), with TP samples generally receiving higher weights, especially in early training. For example, on GMF with ML-1M, TP weights quickly converge near $10^{-1}$, while FP weights stay around $10^{-3}$. These results confirm that our reweighting mechanism effectively distinguishes clean from noisy data and guides learning accordingly.

\subsection{Hyperparameter Sensitivity Analysis (RQ4)}
\label{sec:HyperparmeterAnalyse}

\begin{figure}[!t]
\centering
\begin{subfigure}[t]{0.550\linewidth}
    \centering
    \includegraphics[width=1.0\linewidth, keepaspectratio]{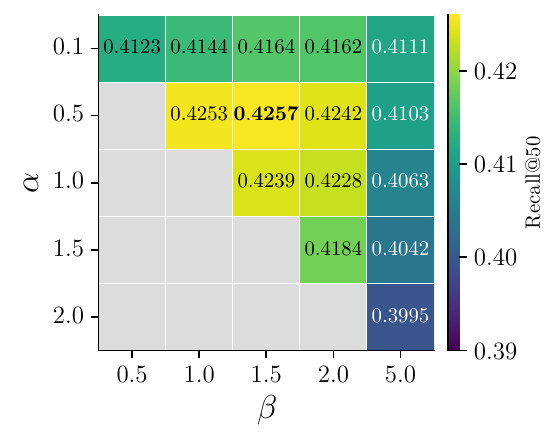}
    \Description{Heatmap showing Recall@50 performance of NeuMF on ML-1M over different values of alpha and beta hyperparameters.}
    \vspace{-1.7em}
    \caption{}
    \label{fig:HyperParameter_NeuMF_Movielens_Rough}
\end{subfigure}
\begin{subfigure}[t]{0.440\linewidth}
    \centering
    \includegraphics[width=1.0\linewidth, keepaspectratio]{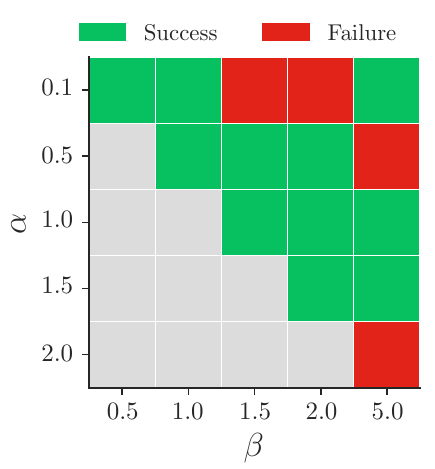}
    \Description{Heatmap of Hessian-based concavity test results for selected hyperparameter points.}
    \vspace{-1.7em}
    \caption{}
    \label{fig:Hessian_valid_result_heatmap}
\end{subfigure}
\vspace{-1.2em}
\caption{(a) Recall@50 performance landscape over $[\alpha, \beta]$ using NeuMF on ML-1M. (b) Hessian-based concavity test at 15 representative points from (a).}
\label{fig:performance_landscape_and_hessian_std}
\end{figure}

We evaluated the sensitivity of EARD to its two key hyperparameters, $\alpha$ and $\beta$, which define the bounds of entity weights. A robust model should retain strong performance under small hyperparameter variations. As shown in Figure~\ref{fig:performance_landscape_and_hessian_std}, the Recall@50 landscape for NeuMF on ML-1M is smooth and unimodal, with a broad region of high performance. This pattern is consistent across all backbones and datasets, as illustrated in Figures~\ref{fig:Hessian_ori_heatmap} and~\ref{fig:all_heat_pic} in the appendix.

To quantitatively support the visual observation of a smooth, unimodal landscape, we performed a local curvature analysis using a discrete Hessian-based concavity test (Algorithm~\ref{alg:hessian_valid_concave} in Appendix~\ref{sec:validConcave}). The results, shown in Figure~\ref{fig:Hessian_valid_result_heatmap}, reveal a strong correlation between mathematically verified concave regions and the high-performance areas of the landscape. This alignment confirms that the objective surface exhibits a favorable and well-structured geometry, devoid of erratic local optima. Such a ``hill-shaped'' topology is particularly valuable, as it suggests that EARD is robust to hyperparameter selection; small perturbations near the optimum do not lead to significant performance degradation. As a result, the framework remains easy to tune in practice.

\subsection{Efficiency Analysis (RQ5)}
\label{sec:efficiencyAnalysis}

In practical deployment, it is essential to balance model performance and computational efficiency. To answer \textbf{RQ5}, we compare the overhead of \textbf{EARD} with the vanilla training pipeline and other denoising methods. We measure peak GPU memory usage and the number of training epochs to convergence. To ensure fair and consistent evaluation, we report the number of epochs instead of wall-clock time, which can be affected by varying server conditions. Table~\ref{tab:mem_and_epochs_cost} presents the results, where validation is conducted after every epoch and early stopping is applied if no improvement is observed for 10 consecutive epochs.

The results demonstrate that EARD offers a clear efficiency advantage. Although it introduces a modest and predictable memory overhead of 5 to 15 percent compared to the vanilla baseline, methods such as \textbf{DeCA} and \textbf{BOD} consume substantially more resources, requiring up to \textbf{3x} and \textbf{2x} more memory, respectively. This indicates that EARD achieves performance improvements without incurring the high memory costs typical of co-training auxiliary models or applying bi-level optimization. Regarding convergence speed, while advanced denoising methods, including ours, DeCA, and BOD, occasionally require more epochs than vanilla training, we attribute this to a more thorough optimization process that avoids premature convergence and yields better final performance.

\begin{table}[htbp]
\centering
\small
\caption{Computational efficiency comparison of denoising methods. Each cell reports peak GPU memory usage (MB, \textbf{Mem}) and training epochs to convergence (\textbf{Ep}). EARD (Ours) generally achieves favorable efficiency with low cost.}
\vspace{-1.0em}
\resizebox{\linewidth}{!}{
\setlength{\tabcolsep}{1.2mm}
\begin{tabular}{@{}lcccccccccccc@{}}
\toprule
\multirow{2}{*}{\textbf{Method}} 
& \multicolumn{6}{c}{\textbf{GMF}} 
& \multicolumn{6}{c}{\textbf{NeuMF}} \\
\cmidrule(lr){2-7} \cmidrule(lr){8-13}
& \multicolumn{2}{c}{ML-1M} & \multicolumn{2}{c}{Yelp} & \multicolumn{2}{c}{Amazon-Book} 
& \multicolumn{2}{c}{ML-1M} & \multicolumn{2}{c}{Yelp} & \multicolumn{2}{c}{Amazon-Book} \\
\cmidrule(lr){2-3} \cmidrule(lr){4-5} \cmidrule(lr){6-7}
\cmidrule(lr){8-9} \cmidrule(lr){10-11} \cmidrule(lr){12-13}
& \textbf{Mem} & \textbf{Ep} & \textbf{Mem} & \textbf{Ep} & \textbf{Mem} & \textbf{Ep}
& \textbf{Mem} & \textbf{Ep} & \textbf{Mem} & \textbf{Ep} & \textbf{Mem} & \textbf{Ep} \\
\midrule
Vanilla 
& 528  & 46  & 1232  & 172  & 1894  & 131 
& 958  & 36  & 6476  & 91   & 10890 & 89 \\
DeCA   
& 1510 & 63  & 4050  & 218  & 7346  & 182 
& 2972 & 49  & 25347 & 123  & 34325 & 132 \\
BOD    
& 1140 & 51  & 2713  & 153  & 4621  & 144 
& 2130 & 41  & 14793 & 84   & 28641 & 101 \\
Ours   
& 559  & 48  & 1330  & 106  & 2124  & 188 
& 1063 & 32  & 7382  & 77   & 11870 & 96 \\
\bottomrule
\end{tabular}
}
\label{tab:mem_and_epochs_cost}
\end{table}

In summary, EARD achieves the favorable balance between performance and efficiency. It delivers state-of-the-art results (Table~\ref{tab:overall_performance}) while maintaining a minimal memory footprint compared to other strong baselines. Its lightweight and highly parallelizable design makes it not only effective but also practical for large-scale industrial recommender systems, where the high memory demands of methods such as DeCA and BOD render them impractical.

\section{Conclusion}
\label{sec:Conclusion}

We introduced EARD, a lightweight and effective framework for denoising implicit feedback by incorporating both entity- and interaction-level weight. By leveraging average loss as a proxy for noise, EARD captures user- and item-specific noise patterns and assigns fine-grained confidence scores through a nonparametric ECDF-based weighting scheme. Extensive experiments across multiple datasets and backbone models demonstrate consistent improvements over state-of-the-art denoising methods, with lower computational cost and minimal hyperparameter tuning. These results highlight the importance of entity-aware modeling and confirm EARD as a practical and scalable solution for recommendation in noisy environments.

\bibliographystyle{ACM-Reference-Format}
\bibliography{sample-base}

\appendix

\section{Hyperparameter Settings}
\label{sec:hyperparameterAnalysis}

We report the optimal hyperparameter for \textbf{EARD} framework, which is controlled by two key parameters, $\alpha$ and $\beta$. These were selected via systematic grid search to ensure stable and high-performing results. To account for variations in data characteristics and sparsity, hyperparameters were tuned independently for each model–dataset pair. All results are based on test set performance under the selected settings. Table~\ref{tab:optimal_hyperparameters} summarizes the final $[\alpha, \beta]$ values, facilitating reproducibility and fair comparison in future studies.

\begin{table}[htbp]
\centering
\setlength{\tabcolsep}{1mm}
\small
\caption{Optimal $[\alpha, \beta]$ values for EARD across backbone models and datasets, selected via grid search.}
\begin{tabular}{@{}lccc@{}}
\toprule
\textbf{Model} & \textbf{ML-1M} & \textbf{Yelp} & \textbf{Amazon-book} \\
\midrule
GMF   & $[1.0, 2.0]$ & $[0.9, 1.0]$  & $[0.14, 0.4]$ \\
NeuMF & $[0.5, 1.5]$ & $[0.05, 0.1]$ & $[0.05, 0.1]$ \\
CDAE  & $[0.5, 1.5]$ & $[0.1, 0.5]$  & $[0.1, 0.5]$  \\
\bottomrule
\end{tabular}
\label{tab:optimal_hyperparameters}
\end{table}

\section{Comparison of Hyperparameter Design}
\label{sec:hyperparameterStat}

Beyond accuracy, practical usability is critical for deploying denoising frameworks in real-world systems. A key factor is the number and sensitivity of hyperparameters. Many state-of-the-art methods, though theoretically sound, introduce numerous non-intuitive hyperparameters that require extensive tuning, increasing computational cost and complicating deployment.

\begin{table}[htbp]
\centering
\small
\caption{Overview of hyperparameter complexity in representative denoising methods. EARD requires only two parameters, offering lower tuning overhead compared to the multi-component designs of DeCA, UDT, and SGDL.}
\vspace{-1.0em}
\setlength{\tabcolsep}{1mm}
\begin{tabular}{@{}lcl@{}}
\toprule
\textbf{Method} & \textbf{\#Params} & \textbf{Hyperparameters and Brief Descriptions} \\
\midrule
\multirow{3}{*}{DeCA} & \multirow{3}{*}{3} & $\alpha$: KL-divergence balance weight. \\
& & $C_+$: Penalize negatives in positive-denoising. \\
& & $C_-$: Penalize positives in negative-denoising. \\
\midrule
\multirow{2}{*}{UDT} & \multirow{2}{*}{4} & $a, b$: Weight mapping range. \\
& & $a', b'$: Temperature mapping range. \\
\midrule
\multirow{5}{*}{SGDL} & \multirow{5}{*}{5} & $h$: Memorization window size. \\
& & $\eta_2$: Weight function learning rate. \\
& & $d_w$: Weight function MLP size. \\
& & $d_l$: Scheduler LSTM size. \\
& & $\tau$: Gumbel-Softmax temperature. \\
\midrule
{Ours} & \textbf{2} & ${\alpha, \beta}$: {Entity weight mapping range.} \\
\bottomrule
\end{tabular}
\label{tab:hyperparameter_comparison_concise}
\end{table}

To highlight EARD’s practical advantages, Table~\ref{tab:hyperparameter_comparison_concise} compares its hyperparameter design with several strong baselines. Methods such as DeCA, UDT, and SGDL rely on complex components such as KL-divergence balancing, temperature scheduling, and meta-learning modules, each demanding careful tuning. In contrast, EARD introduces only two intuitive hyperparameters, $\alpha$ and $\beta$, whose impact on performance is both stable and predictable. Empirical analysis reveals a smooth concave performance landscape, as shown in Figure~\ref{fig:performance_landscape_and_hessian_std} and Appendix Figures~\ref{fig:Hessian_ori_heatmap} and \ref{fig:all_heat_pic}, which facilitates efficient tuning and helps avoid poor local optima. This combination of low complexity and high tunability reduces engineering overhead and improves robustness in practical deployment.

\section{Base Weighting Strategies Analysis}
\label{sec:base_weight_computation_methods}





\subsection{ECDF-based Weighting}

\begin{algorithm}[htbp]
\small
\caption{ECDF-based Weighting (Our Method)}
\label{alg:ecdf}
\begin{algorithmic}[1]
\REQUIRE 
\STATE \textbf{Input:} Set of training instances $\mathcal{D}^{(t)} = \{ (u, i, y_{ui}) \}$ for epoch $t$.
\STATE \textbf{Input:} Corresponding loss values $L = \{ \ell_{ui} \mid (u, i, y_{ui}) \in \mathcal{D}^{(t)} \}$.
\ENSURE \textbf{Output:} Confidence weights $W = \{ w_{ui} \mid (u, i, y_{ui}) \in \mathcal{D}^{(t)} \}$ for each instance.

\STATE $n \leftarrow |L|$ \COMMENT{Total number of training instances}
\STATE $\boldsymbol{\ell}_{\text{valid}} \leftarrow \{ -\ell_{ui} \mid \ell_{ui} \in L \}$ 
\STATE $\boldsymbol{\ell}_{\text{sorted}} \leftarrow \operatorname{Sort}(\boldsymbol{\ell}_{\text{valid}})$ 
\STATE $\mathbf{r} \leftarrow \text{SearchSorted}(\boldsymbol{\ell}_{\text{sorted}}, \boldsymbol{\ell}_{\text{valid}})$ \COMMENT{Get rank (starting from 1) for each value in $\boldsymbol{\ell}_{\text{valid}}$ within the sorted list}
\STATE \textbf{For each} instance $(u, i)$ \textbf{with loss} $\ell_{ui}$:
\STATE \quad $w_{ui\text{,base}} \leftarrow (r_{ui} - 0.5) / n$ \COMMENT{using Hazen plotting position}
\STATE \textbf{return} $W \leftarrow \{ w_{ui\text{,base}} \mid (u, i, y_{ui}) \in \mathcal{D}^{(t)} \}$
\end{algorithmic}
\end{algorithm}

\subsection{Gaussian Mixture Model (GMM)}


\begin{algorithm}[htbp]
\small
\caption{Gaussian Mixture Model (GMM) Weighting}
\label{alg:gmm}
\begin{algorithmic}[1]
\REQUIRE 
\STATE \textbf{Input:} Set of training instances $\mathcal{D}^{(t)} = \{ (u, i, y_{ui}) \}$ for epoch $t$.
\STATE \textbf{Input:} Corresponding loss values $L = \{ \ell_{ui} \mid (u, i, y_{ui}) \in \mathcal{D}^{(t)} \}$.
\ENSURE \textbf{Output:} Confidence weights $W = \{ w_{ui} \mid (u, i, y_{ui}) \in \mathcal{D}^{(t)} \}$ .

\STATE $n \leftarrow L$
\COMMENT{Total number of training instances in the current epoch}

\STATE $\boldsymbol{\ell}_{\text{neg}} \leftarrow \{ -\ell_{ui} \mid \ell_{ui} \in L \}$ \COMMENT{Critical: Negate losses. Higher value in $\ell_{\text{neg}}$ indicates cleaner sample}
\STATE \textbf{Fit} a 2-component Gaussian Mixture Model (GMM) on the set $\boldsymbol{\ell}_{\text{neg}}$.
\STATE Let the two fitted Gaussian components be $k=1, 2$ with means $\mu_1, \mu_2$.
\STATE $k_{\text{clean}} \leftarrow \arg\max_{k \in \{1,2\}} (\mu_k)$ \COMMENT{Identify the component with the larger mean as the 'clean' component}
\STATE \textbf{For each} value $\ell_{\text{neg}}^{ui}$ in $\boldsymbol{\ell}_{\text{neg}}$ \COMMENT{Iterate over the set of negated losses}
\STATE \quad $p_{ui} \leftarrow \text{PosteriorProbability}(\ell_{\text{neg}}^{ui} \text{ belongs to component } k_{\text{clean}})$ 
\STATE \quad $w_{ui} \leftarrow p_{ui}$ 
\STATE \textbf{return} $W \leftarrow \{ w_{ui} \mid (u, i, y_{ui}) \in \mathcal{D}^{(t)} \}$ 
\end{algorithmic}
\end{algorithm}

\subsection{\texorpdfstring{Hard Thresholding (Top-$k$ Selection)}{Hard Thresholding (Top-k Selection)}}

Hard Thresholding (Top‑$k$ Selection) is a denoising method based on the small-loss assumption: lower-loss samples are more likely clean. Each epoch, instances are ranked by loss, and the top $\rho$ fraction (e.g., $\rho = 0.8$) are assigned weight 1; the rest receive 0. This binary weighting makes it a hard selection scheme.

\subsection{Linear Scaling}
Linear Scaling is a soft-weighting method that assigns continuous weights based on loss values. Losses are linearly mapped to the range $[0, 1]$, with the minimum loss receiving weight 1 and the maximum 0. While this yields smooth weight assignments, the method is sensitive to outliers, which can distort the scaling and suppress informative samples.

\section{Noise is Not Random: An Empirical Study of Entity-Level Bias}
\label{sec:entityNoiseRateStat}

\begin{figure}[htbp]
\centering
\includegraphics[width=1.0\linewidth]{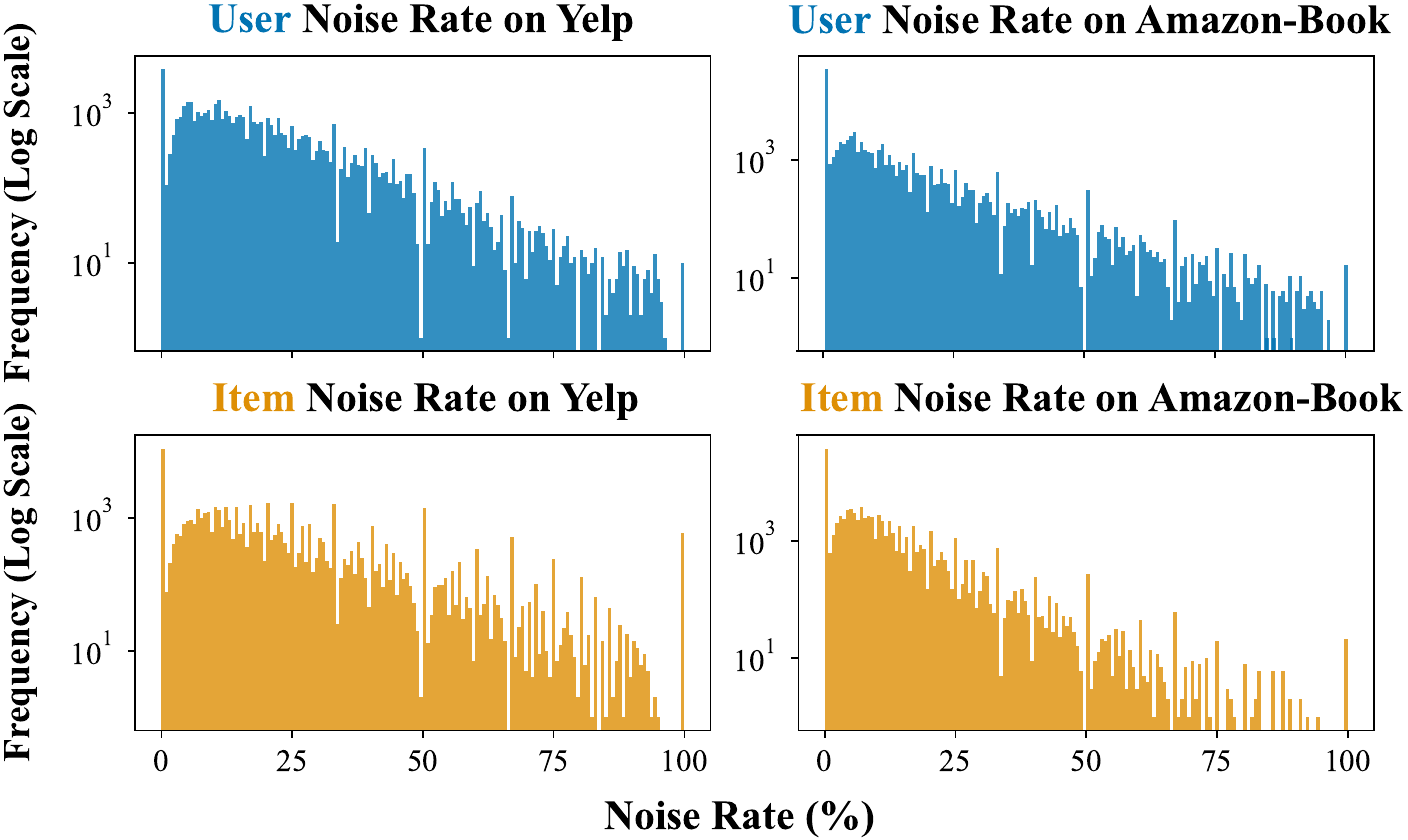}
\Description{Histograms of user and item noise rates on the ML-1M, Yelp, and Amazon-Book datasets showing the proportion of interactions with ratings <=3, with y-axis on a logarithmic scale.}
\caption{Histograms of user (blue) and item (yellow) noise rates on the ML-1M, Yelp, and Amazon-Book datasets.
}
\label{fig:all_noisy_rate_distribution}
\end{figure}

We empirically find that noise in implicit feedback is not random but is systematically linked to the user and item entities involved. A key commonality across all evaluated datasets (ML-1M, Yelp, and Amazon-Book) is the broad and non-uniform distribution of entity-level noise. We define an entity's noise rate as the proportion of its interactions with ratings at or below 3. As shown in Figures~\ref{fig:all_noisy_rate_distribution} and~\ref{fig:ml_1m_noisy_rate_distribution}, users and items are spread across the entire 0\% to 100\% noise spectrum in all datasets.

Crucially, a non-negligible number of entities exists within each segment of the noise distribution. This indicates that both highly reliable (low-noise) and highly unreliable (high-noise) entities are consistently present, regardless of the dataset's specific characteristics. This widespread heterogeneity invalidates the assumption that all entities are equally trustworthy and provides a strong motivation for our proposed EARD framework, which adaptively estimates user reliability and item reputation for more robust learning.

\section{Analysis of Loss Distribution Diversity Across Models and Datasets}
\label{sec:lossDistributionStat}

\begin{figure}[htbp]
\centering
\includegraphics[width=1.0\linewidth]{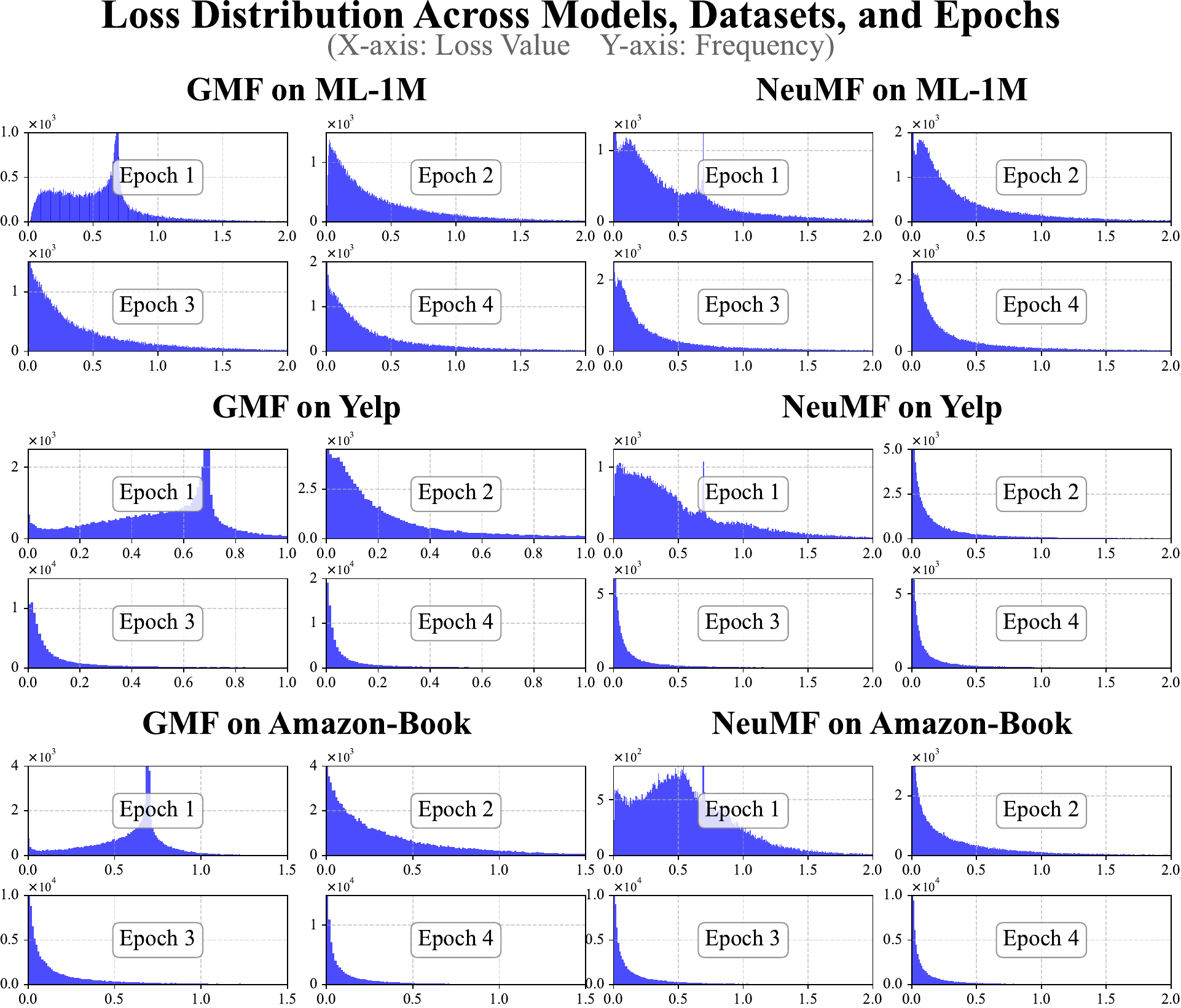}
\Description{Histograms of training loss distributions for GMF, NeuMF, and CDAE models on ML-1M, Yelp, and Amazon-Book datasets over the first four epochs.} 
\caption{Histograms of training loss distributions over the first four epochs for GMF and NeuMF on the ML-1M, Yelp, and Amazon-Book datasets.}
\label{fig:loss_distribution_stat}
\end{figure}

Figure~\ref{fig:loss_distribution_stat} shows the loss distributions over the first four training epochs for two representative recommendation models, GMF and NeuMF, on the ML-1M, Yelp, and Amazon-Book datasets. Each subplot presents a histogram of instance-level training losses at a specific epoch, where the x-axis indicates the loss value and the y-axis shows the sample frequency.

As observed, the loss distributions exhibit significant heterogeneity. They vary considerably not only across different models and datasets but also evolve dynamically across training epochs. This high degree of variability makes it unrealistic to presuppose any single, fixed distributional form for the losses. Instead, our empirical results show that the distributions for GMF and NeuMF are often complex and non-standard, frequently exhibiting long-tailed and heavily skewed shapes. These distributions also contain numerous outliers, further complicating any attempt at parametric modeling.

These empirical findings highlight the critical need for a base weighting strategy that is both distribution-agnostic and robust to outliers. Parametric methods like Gaussian Mixture Models (GMM) or threshold-based heuristics such as Top-k Selection rely on strong distributional assumptions (e.g., unimodality or Gaussianity), which are clearly violated by the observed diverse and skewed patterns. In contrast, a rank-based approach like ECDF avoids such restrictive assumptions. It naturally resists the influence of extreme values by operating on relative ranks rather than absolute magnitudes, thereby providing stable and consistent weight allocation across diverse training conditions. This strongly motivates our use of ECDF for noise-aware reweighting.

\section{Analyzing the Correlation Between Loss Patterns and Noise Rates}
\label{sec:avgLoss_noiseRate}

\begin{figure}[htbp]
\centering
\begin{subfigure}[t]{1.0\linewidth}
\centering
\includegraphics[width=1.0\linewidth]{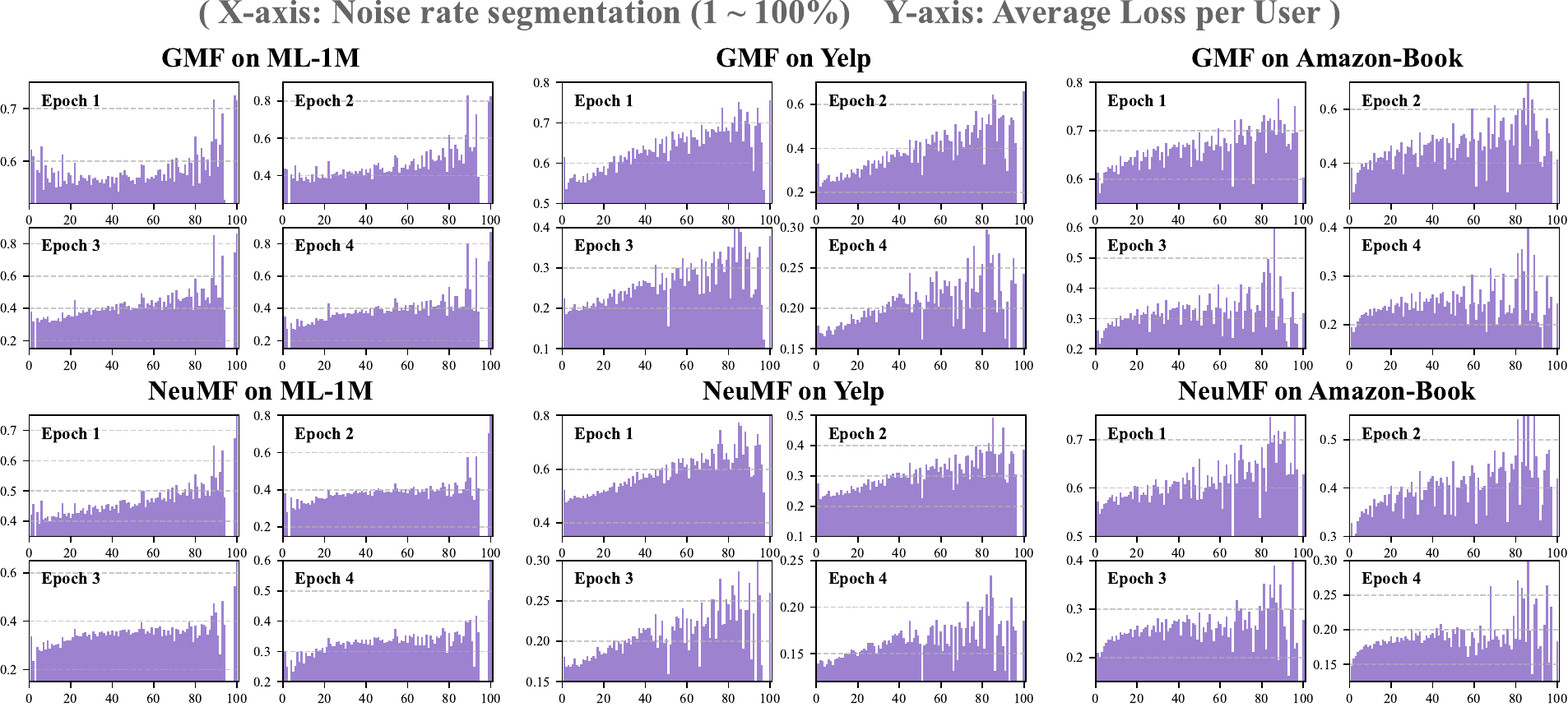}
\Description{Line plots showing average per-user loss across 100 noise rate percentiles over four epochs for GMF and NeuMF on ML-1M, Yelp, and Amazon-book datasets.}
\caption{}
\label{fig:user_loss_histograms}
\end{subfigure}
\begin{subfigure}[t]{1.0\linewidth}
\centering
\includegraphics[width=1.0\linewidth]{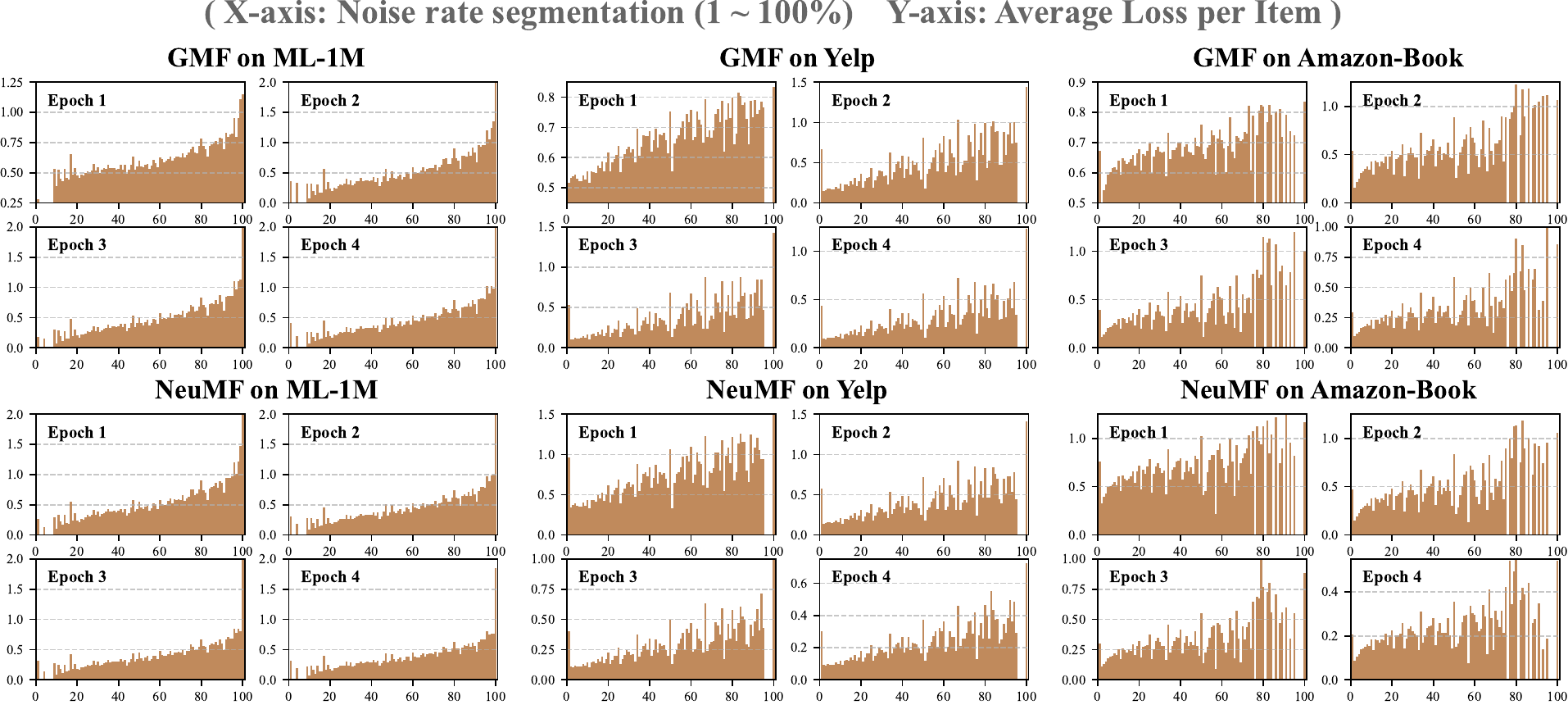}
\Description{Line plots showing average per-item loss across noise rate percentiles over four epochs for GMF and NeuMF on the same datasets.}
\caption{}
\label{fig:item_loss_histograms}
\end{subfigure}
\caption{Loss patterns segmented by entity noise rate. (a) Average per-user loss and (b) average per-item loss across 100 noise rate percentiles. A higher entity-level noise rate generally correlates with a higher average training loss, validating its use as a practical reliability proxy.}
\label{fig:item_user_loss_histograms}
\end{figure}

Figure~\ref{fig:item_user_loss_histograms} presents a noise-aligned analysis of model behavior from both user and item perspectives. Training instances are grouped by the noise rates of their associated interactions (ranging from 1\% to 100\%), and average losses are computed within each percentile group over the first four training epochs.

Figure~\ref{fig:user_loss_histograms} illustrates the per-user average loss distributions for GMF and NeuMF across the ML-1M, Yelp, and Amazon-Book datasets. In all cases, a clear and approximately monotonic trend emerges: as user noise rates increase, their corresponding average losses also rise. For example, under GMF on Yelp at Epoch 4, users in the top 20\% noise group have losses between 0.6 and 0.8, while those in the bottom 20\% remain around 0.2 to 0.3. This trend suggests that user-level loss serves as a meaningful indicator of reliability.

Similarly, Figure~\ref{fig:item_loss_histograms} shows consistent patterns at the item level. Items in higher noise rates tend to incur higher average losses, reinforcing the use of loss as a proxy for identifying noisy interactions.

These results empirically support the core intuition behind EARD: entity-level loss serves as a reliable and quantifiable signal of interaction trustworthiness. The clear alignment between loss and noise across users and items validates loss-based metrics for adaptive reweighting and denoising.

\section{Analysis of Hyperparameter Sensitivity}
\label{sec:validConcave}

We analyze the impact of the weighting hyperparameters $\alpha$ and $\beta$ on model performance and examine the geometry of the objective surface induced by EARD. Our study spans multiple models and datasets to evaluate both local and global landscape properties.

\subsection{Generalization Across Models and Datasets}

\begin{figure}[htbp]
\centering
\includegraphics[width=1.0\linewidth, keepaspectratio]{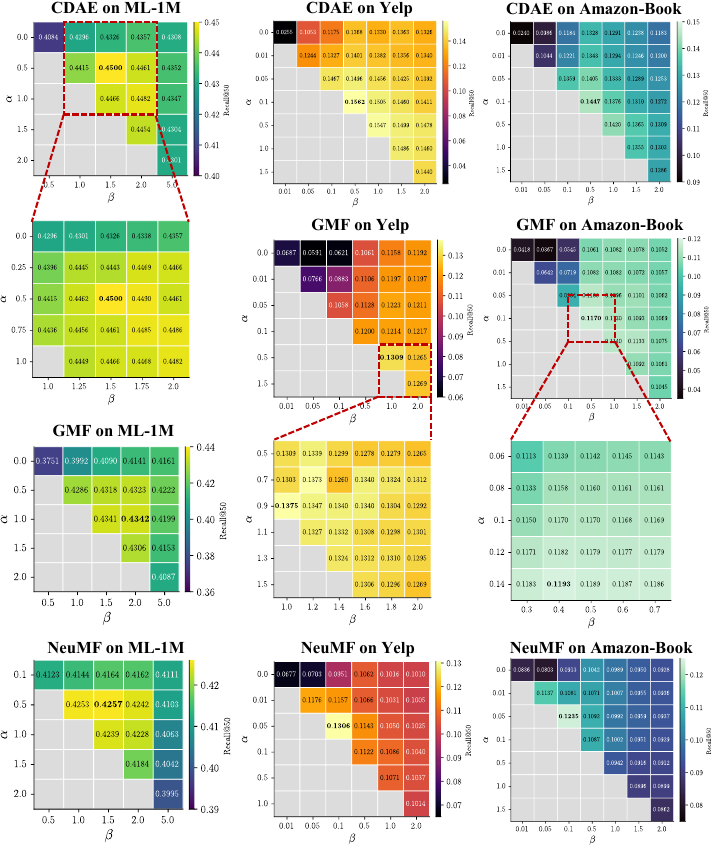}
\Description{Nine heatmaps showing Recall@50 values across alpha and beta hyperparameter combinations for GMF, NeuMF, and CDAE models on ML-1M, Yelp, and Amazon-book datasets. Brighter colors indicate higher performance, with most models showing a concave performance landscape.}
\caption{Hyperparameter sensitivity analysis across all models and datasets. The heatmaps of Recall@50 performance for varying $\alpha$ and $\beta$ generally reveal a smooth, approximately concave landscape, validating the framework's robustness to its hyperparameter choices.}
\label{fig:all_heat_pic}
\end{figure}

To examine whether the approximate concavity of the objective surface generalizes across model architectures and datasets, we plot Recall@50 heatmaps for GMF, NeuMF, and CDAE on three benchmarks: ML-1M, Yelp, and Amazon-Book (Figure~\ref{fig:all_heat_pic}). Each heatmap visualizes the average Recall@50 over a grid of $(\alpha, \beta)$ values.

Across all settings, the heatmaps generally reveal hill-like patterns. This global concavity is observed in both simple (e.g., GMF) and expressive (e.g., NeuMF) models, and remains stable across datasets with varying levels of sparsity and noise. These results support the hypothesis that the EARD objective surface is approximately concave in practice, with a well-defined local optimum. This property facilitates stable optimization using gradient-based or local search methods, and highlights the generalizability of the proposed hyperparameter design.

\subsection{Concavity Verification via Hessian Matrix}

\begin{figure}[htbp]
\centering
\includegraphics[width=1.0\linewidth]{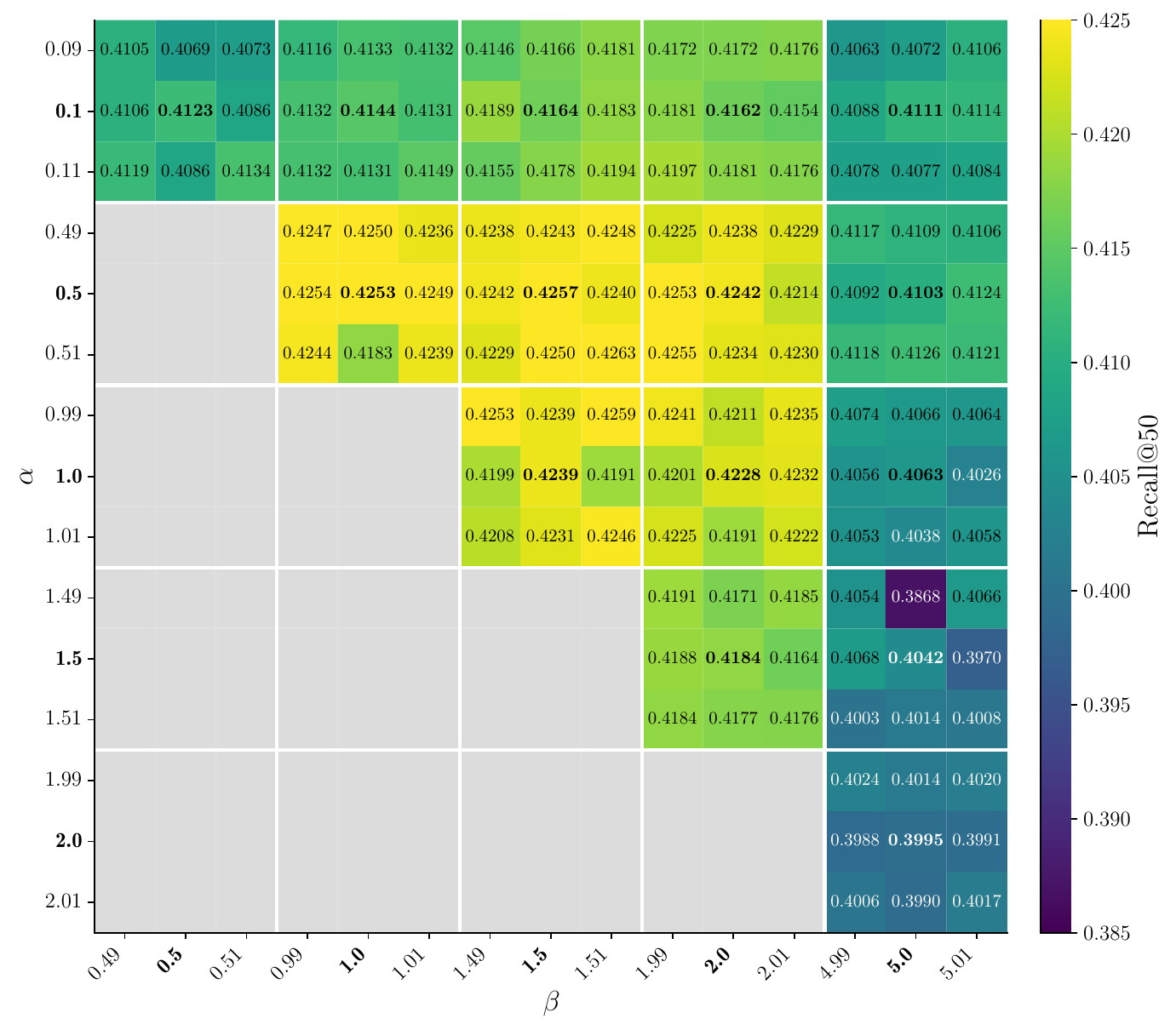}
\Description{Heatmap showing Recall@50 scores of the NeuMF model on ML-1M across a fine grid of alpha and beta hyperparameter values. Brighter areas indicate higher performance.}
\caption{Heatmap of NeuMF performance on ML-1M (Recall@50), generated by finely sampling $[\alpha, \beta]$ around 15 anchor points (bolded). Brighter regions indicate better performance and align with Hessian-based analysis, empirically supporting local concavity in high-performance areas.}
\label{fig:Hessian_ori_heatmap}
\end{figure}

We adopt a second-order approximation method to examine the local curvature of the performance landscape around specific $(\alpha, \beta)$ configurations. Specifically, we calculate the discrete Hessian matrix over a $3 \times 3$ grid of Recall@50 scores centered at each anchor point. As outlined in Algorithm~\ref{alg:hessian_valid_concave}, the second-order partial derivatives are computed using finite difference methods, and the determinant of the Hessian matrix $\det(H)$ is used to assess local curvature. A point is considered locally concave if $H_{11} < 0$ and $\det(H) > 0$, indicating a ``hill''-shaped performance landscape.

To assess the smoothness and concavity of the performance surface, we conduct a fine-grained sweep of $(\alpha, \beta)$ hyperparameters for the NeuMF model on ML-1M, as shown in Figure~\ref{fig:Hessian_ori_heatmap}. Sampling around 15 anchor points at a 0.01 resolution reveals a clear unimodal landscape with peak performance at the center and gradual decline toward the edges. This pattern aligns with the Hessian-based results in Figure~\ref{fig:Hessian_valid_result_heatmap}, where most points exhibit local concavity. These results further confirm the unimodal geometry of the objective surface and demonstrate that EARD maintains stable performance under small hyperparameter variations.

\noindent \textbf{Definition:} \textit{The function $f(x, y)$ denotes the model's evaluation score (e.g., Recall@50) on the test set, obtained after training with hyperparameters $\alpha = x$ and $\beta = y$.}

\begin{algorithm}[htbp]
\small
\caption{Verify Local Concavity via Discrete Hessian Test}
\label{alg:hessian_valid_concave}
\begin{algorithmic}[1]
\REQUIRE A $3 \times 3$ grid of function values $\mathbf{F}$ centered at $f_c = f(x, y)$, with step sizes $h_x, h_y$.
\ENSURE Local curvature: Concave, Convex, Saddle Point, or Indeterminate.

\STATE \COMMENT{Compute second-order partial derivatives using finite differences from grid $\mathbf{F}$}
\STATE $H_{11} \gets (F_{x+h, y} - 2f_c + F_{x-h, y}) / h_x^2$
\STATE $H_{22} \gets (F_{x, y+h} - 2f_c + F_{x, y-h}) / h_y^2$
\STATE $H_{12} \gets (F_{x+h, y+h} - F_{x-h, y+h} - F_{x+h, y-h} + F_{x-h, y-h}) / (4 h_x h_y)$

\STATE Compute determinant: $\det(H) \gets H_{11} H_{22} - H_{12}^2$.

\IF{$H_{11} < 0$ and $\det(H) > 0$}
    \RETURN \textbf{Locally Concave} (`hill')
\ELSIF{$H_{11} > 0$ and $\det(H) > 0$}
    \RETURN \textbf{Locally Convex} (`basin')
\ELSIF{$\det(H) < 0$}
    \RETURN \textbf{Saddle Point}
\ELSE
    \RETURN \textbf{Indeterminate}
\ENDIF
\end{algorithmic}
\end{algorithm}

\end{document}